\documentstyle{article}
\def\be{\begin{equation}}
\def\ee{\end{equation}}
\def\bea{\begin{eqnarray}}
\def\eea{\end{eqnarray}}
\def\l{\label}
\def\ct{\cite}
\def\r{\ref}

\def\lgth{[\,\mbox{length}\,]}
\def\Th{\Theta}
\def\sig{\sigma}
\def\om{\omega}
\def\ndot{\dot{n}}
\def\4pig{\sfrac{4\pi G}{c^{4}}}
\def\gam{\gamma}
\def\d{\delta}
\def\eps{\epsilon}
\def\D{\mbox{D}}
\def\lgl{\langle}
\def\rgl{\rangle}
\def\tl{\tilde}
\def\hsp5{\hspace{5mm}}
\newcommand{\sfrac}[2]{{\textstyle{#1\over#2}}}
\def\case#1/#2{\textstyle\frac{#1}{#2}}
\def\cqg{{\em Class. Quantum Grav.\/} }
\def\grg{{\em Gen. Rel. Grav.\/} }
\def\prd{{\em Phys. Rev.\/} D }
\def\prl{{\em Phys. Rev. Lett.\/} }
\newcommand{\enl}{\\\hfill\rule{0pt}{0pt}}
\title{\sc Quasi-Newtonian dust cosmologies}
\author{{\sc Henk van Elst\thanks{e-mail:
henk@gmunu.mth.uct.ac.za}
\ \& George F R Ellis\thanks{e-mail: ellis@maths.uct.ac.za}}\\
{\small\em Department of Mathematics and Applied Mathematics,
University of Cape Town, Rondebosch 7701}\\
{\small\em Cape Town, South Africa}}
\date{\normalsize{May 6, 1998}}
\begin{document}
\sloppy
\maketitle
\begin{abstract}
Exact dynamical equations for a generic dust matter source field in
a cosmological context are formulated with respect to a
non-comoving Newtonian-like timelike reference congruence and
investigated for internal consistency. On the basis of a lapse
function $N$ (the relativistic acceleration scalar potential) which
evolves along the reference congruence according to $\dot{N} =
\alpha\,\Theta\,N$ ($\alpha = \mbox{const}$), we find that
consistency of the quasi-Newtonian dynamical equations is not
attained at the first derivative level. We then proceed to show
that a self-consistent set can be obtained by linearising the
dynamical equations about a (non-comoving) FLRW background. In this
case, on properly accounting for the first-order momentum density
relating to the non-relativistic peculiar motion of the matter,
additional source terms arise in the evolution and constraint
equations describing small-amplitude energy density fluctuations
that do not appear in similar gravitational instability scenarios
in the standard literature.

\end{abstract}
\centerline{\bigskip\noindent PACS number(s): 04.20.-q, 98.80.Hw, 
95.30.Sf, 98.65.Dx}
\newpage

\section{Introduction}
The aim of this paper is to investigate the Newtonian limit of the
Theory of General Relativity in a cosmological context, when
conditions relating to gravitational physics are significantly
different from the more traditional quasi-stationary and
asymptotically flat situations where the Newtonian and
post-Newtonian limits can usually be derived and are subject to
precise experimental tests. The importance of this is that many
astrophysical calculations on the formation of large-scale
structure in the Universe rely on such a limit, because most of
these calculations are done in a Newtonian way. A number of
derivations of such a limit are given in the literature, see, e.g.,
Refs. \ct{pee80}, \ct{ber92} and \ct{holwal97}; indeed, there is a
viewpoint that cosmology is essentially a Newtonian affair, with
the relativistic theory only needed for examination of some
observational relations. However:

(A) The true classical theory of gravitation is General Relativity;
Newtonian theory is only a good theory of gravitation when it is a
good approximation to the results obtained from General Relativity;

(B) We have to extend standard Newtonian theory (which strictly can
only deal with quasi-stationary isolated systems in an
asymptotically flat spacetime) in some way or another to deal with
a non-stationary cosmological context. This extension needs to be
clearly spelt out.

Significant questions remain in regard to both points. These
include:

(1) Full General Relativity theory involves {\em ten gravitational
potentials\/} (combined in a tensorial variable), subject to the
ten Einstein field equations (`EFE'), but Newtonian theory involves
only one (a scalar variable), subject to the one Poisson field
equation (`PFE'); how does it arise that the other nine potentials
and equations can be ignored in the Newtonian limit? Given that
nine equations of the full theory are not satisfied even in some
limiting sense, how do we know when Newtonian cosmological
solutions correspond to consistent relativistic solutions of the
full set of equations? Part of the answer is that the coordinate
freedom of General Relativity accounts for four of these
potentials; however, this still leaves another five to account for,
and we have examples of Newtonian solutions with {\em no\/} General
Relativity analogues \ct{ell97,ellhve98}. Consequently we need to
be concerned as to how well standard Newtonian theory represents
the results of General Relativity in the cosmological context we
have in mind, when we take a `Newtonian limit'.

(2) The issue of {\em boundary conditions\/} for Newtonian theory
in the cosmological context is problematic even in the context of
exactly spatially homogeneous (and spatially isotropic)
cosmological models \ct{hecsch56}; no fully adequate theory exists
in the more realistic almost-homogeneous case. Numerical
simulations, for example, usually rely on periodic boundary
conditions, which correspond to the real Universe only if we live
in a `small universe' \ct{ellsch86} in which there is a
long-wavelength cut-off in the spectrum of inhomogeneities of its
large-scale structure. Analytic solutions usually rely on
asymptotically flat conditions which are manifestly not true in a
realistic, almost-Friedmann--Lema\^{\i}tre--Robertson--Walker
(`FLRW') situation (they implicitly or explicitly assume that
inhomogeneities die off sufficiently far away from the region of
interest).\footnote{Ideally one would impose conditions at a finite
distance rather than infinity \ct{ell84}; however, this `finite
infinity' approach has not yet been developed fully.}

(3) How do we attain a {\em unique propagation equation\/} for the
gravitational scalar potential in a Newtonian cosmology, when
Newtonian theory proper has no such equation? In standard Newtonian
theory, which aims to describe only quasi-stationary settings, this
results from the assumed asymptotically flat condition --- which
does not hold in the cosmological context.

(4) How do we satisfactorily handle the {\em gauge-dependence\/}
that underlies most derivations of a Newtonian limit? Equivalently,
most derivations of the Newtonian limit are highly coordinate
dependent, despite the strongly proclaimed doctrine of covariance
of General Relativity; do we need to revise our view on
this?\footnote{See, e.g., the discussion by Ellis and Matravers
\ct{ellmat95}.}

One approach to these issues is the `frame theory' suggested by
Ehlers \ct {ehl81,ehl97}, which has the advantage of giving precise
theorems on some Newtonian limits, but has the disadvantage of
being more complex than General Relativity. Here we seek a more
direct approach to the Newtonian limit in a cosmological situation,
that is more readily related to the approaches taken in the
astrophysical context. We do so by introducing a class of exact
relativistic cosmologies that we can justifiably call {\em
quasi-Newtonian\/} because of the properties we discuss below, and
which include as special cases linearised models used in many
astrophysical studies, for example in gravitational lensing
theory.\footnote{They are related to the spacetimes given this name
by Bardeen \ct{bar80} when we linearise our models about a
non-comoving FLRW geometry.} In studying these spacetimes, we will
come across each of the above issues, in a context where we can
pose --- but not often solve --- precise questions concerning exact
quasi-Newtonian solutions of the EFE; consequently we will present
some unanswered {\em `Problems'\/} in what follows --- well-posed
questions that should be answerable, but are of such complexity
that we have been able to make only little progress in their
solution. We then examine the effect of linearising about a ({\em
non\/}-comoving) FLRW model, when these questions become tractable
and interesting issues occur in relation to the approaches usually
taken.

In order to limit the complexity of the problem, we will only
consider cosmological models in which the source of spacetime
curvature is given by a pressure-free matter field (`dust'); that
is, we will examine here only the pure gravitational problem. A
further feature of our discussion will be the absence of
gravitational radiation; on the one hand, this is a property of
gravitational fields unknown to the Newtonian theory, and on the
other, gravitational radiation is commonly thought of as not being
of immediate dynamical relevance in structure formation
processes. We proceed as follows: in section \r{sec:qn} we develop
our concept of a quasi-Newtonian dust cosmology in $1+3$ covariant
terms and touch upon issues of the uniqueness of the non-comoving
time slicing chosen. In section \r{sec:13covqn} we present the
relevant $1+3$ covariant dynamical equations. These are derived
from the Ricci and Bianchi identities on using the EFE to
algebraically relate the Ricci curvature to the dust matter source
field. We then fix the time-reparameterisation freedom of the EFE
by proposing to evolve the lapse function along the timelike
reference congruence in a way which includes various choices made
in the literature. In section \r{sec:qncons}, with the given choice
of lapse function evolution, we pursue a consistency analysis of
the dynamical equations to the first derivative level; no
consistency is attained at this stage. A number of dynamically
specialised subcases are then highlighted in section
\r{sec:qnconsres}. In section \r{sec:qnflrwlin} we investigate the
effect of consistently linearising our dynamical equations about a
non-comoving FLRW background. Compared to results in the standard
literature, our approach, which fully accounts for the first-order
momentum density of non-relativistically moving matter, leads to
modifications in the set of evolution and constraint equations. Our
conclusions are contained in section \r{sec:concl}. Useful
computational details such as commutation relations, the set of
$1+3$ covariant dynamical equations cast into an associated $1+3$
orthonormal frame form, and the linearised EFE for a
scalar-perturbed, spatially flat FLRW model have been gathered in
an appendix. Throughout our work we will employ geometrised units
characterised by $c = 1 = 8\pi G/c^{2}$. Some of our notation is
inspired by Ref. \ct{maa97}.

Our discussion concentrates on two main issues: (i) the internal
consistency of the set of {\em exact\/} equations defining
relativistic quasi-Newtonian dust cosmologies; in looking at this,
we need to ask whether or not the time slicing condition we impose
leads to a {\em uniquely\/} defined foliation ${\cal T}$:
$\left\{t=\mbox{const}\right\}$, and have to face the problem of
time-reparameterisation freedom of the dynamical equations of
General Relativity (the EFE); and (ii) how, if at all, do the {\em
exact\/} equations reduce under FLRW-linearisation to the Newtonian
form encountered in many works on gravitational instability of
matter in non-relativistic motion, i.e., what happens to the
constraints in this limit, and when is the set of equations so
obtained both consistent and linearisation-stable? Of most direct
relevance to an astrophysically oriented reader is the discussion
given in section \r{sec:qnflrwlin}, where, in relating our approach
to some of those given in the astrophysical literature, we show
that extra terms not always included seem to be needed if the
Newtonian-like equations used are to be consistently derived from
General Relativity. The reader may directly address this material
after becoming acquainted with the definitions and relations
provided in sections \r{sec:qn} and \r {sec:13covqn}.

\section{Quasi-Newtonian cosmologies}
\l{sec:qn}
\subsection{Definition}
\l{subsec:qndef}
We define relativistic {\em quasi-Newtonian\/} cosmologies to be
spacetime manifolds $\left(\,{\cal M},\,{\bf g}\,\right)$, subject
to the EFE, which contain a congruence of worldlines that is {\em
irrotational\/} and {\em shearfree\/}. We call such a congruence a
{\em Newtonian-like timelike congruence\/}, and denote the
associated future-directed, normalised, tangent vector field by
${\bf n}$ ($n_{a}\,n^{a} = -\,1$), the cosmologies by
$\left(\,{\cal M},\,{\bf g},\,{\bf n}\,\right)$.

Most cosmologies will {\em not\/} admit such a congruence of
Newtonian-like worldlines, and, hence, are not quasi-Newtonian;
examples that {\em are\/}, are the FLRW cosmologies.\footnote{And,
in the non-cosmological context, the interior and exterior
Schwarzschild spacetimes.} If a cosmology {\em does\/} admit such a
congruence, it will almost always be {\em unique\/} because it has
a special relation to the Weyl curvature of $\left(\,{\cal
M},\,{\bf g}\,\right)$; we explore this further below. Because it
is irrotational, this congruence is normal to a preferred family of
spacelike 3-surfaces of constant time, ${\cal T}$:
$\left\{t=\mbox{const}\right\}$, corresponding to the spacelike
3-surfaces of constant absolute time in a Newtonian
spacetime. Because of the shearfree condition, these 3-surfaces are
mapped conformally onto each other\footnote{They are not mapped
{\em isometrically\/} onto each other, primarily because we allow
for an expansion of their normals. One could make a case that {\em
Strictly Newtonian-Like\/} solutions should demand this stronger
condition; however, that would exclude most of the expanding
cosmological models that are a prime concern of this paper (but
also are not strictly Newtonian).  Accordingly, we stick to the
weaker definition given above.} by the preferred timelike curves
${\bf n}$. The matter in such a setting will almost always {\em
not\/} be comoving with these curves (or, equivalently, will not be
at rest in the preferred spacelike 3-surfaces). The relativistic
acceleration of these curves can be derived from a scalar potential
that corresponds to the Newtonian gravitational scalar potential.

In more detail: we aim to describe the physics of a pressure-free
matter field from the perspective of a Newtonian-like class of
Eulerian observers comoving with ${\bf n}$. To this end, we apply a
$1+3$ covariant treatment adapted to ${\bf n}$
(\,cf. Refs. \ct{ehl61,ell71,hve96,maa97}\,) rather than the
essentially equivalent ADM $3+1$ formulation of the dynamics of
relativistic spacetime geometries \ct{adm62,yor79}. The family of
spacelike 3-surfaces ${\cal T}$: $\left\{t=\mbox{const}\right\}$
orthogonal to the timelike reference congruence constitutes the
rest 3-spaces of the Eulerian observers, with (time-dependent)
3-metric $h_{ab} := g_{ab} + n_{a}\,n_{b}$ satisfying
$\dot{h}_{\lgl ab\rgl} = 0 = \D_{a}h_{bc}$. As to notation, for any
geometrically defined variable ${\bf T}$, we write
$\dot{T}^{a\dots}{}_{b\dots} :=
n^{c}\nabla_{c}T^{a\dots}{}_{b\dots}$ and
$\D_{a}T^{b\dots}{}_{c\dots} := h^{d}\!_{a}\,h^{b}\!_{e}\dots
h^{f}\!_{c}\dots \nabla_{d}T^{e\dots}{}_{f\dots}$, while angle
brackets stand for a fully symmetric tracefree ${\bf n}$-orthogonal
projection as in, e.g., $\dot{h}_{\lgl ab\rgl} = [\
h^{c}\!_{(a}\,h^{d}\!_{b)} - \sfrac{1}{3}\,h_{ab}\,h^{cd} \
]\,\dot{h}_{cd}$. The timelike normals ${\bf n}$ to the ${\cal T}$:
$\left\{t=\mbox{const}\right\}$ are irrotational:
\be
\l{zerovor}
\om^{a}({\bf n}) = 0 \ ,
\ee
and shearfree:
\be
\l{zerosig}
\sig_{ab}({\bf n}) = 0 \ .
\ee
The assumption (\r{zerosig}) corresponds to the vanishing of the
anisotropic part of the extrinsic curvature of the orthogonal
spacelike 3-surfaces ${\cal T}$: $\left\{t=\mbox{const}\right\}$
(\,these 3-surfaces exist because of property (\r{zerovor})\,). The
setting so outlined shares the features of the `zero-shear
hypersurfaces' gauge conditions introduced by Bardeen in his
well-known work on linearised gauge-invariant cosmological
perturbations \ct{bar80}.\footnote{The normals to the ${\cal T}$:
$\left\{t=\mbox{const}\right\}$ are also non-shearing for Bardeen's
`longitudinal' gauge conditions \ct{bar80}.}

\subsubsection{Zero magnetic Weyl curvature}
Through the $1+3$ covariant Ricci identities for ${\bf n}$
\ct{ell71,hve96}, the kinematical restrictions given by
Eqs. (\r{zerovor}) and (\r{zerosig}) immediately lead to zero {\em
magnetic\/} Weyl curvature of $\left(\,{\cal M},\,{\bf g},\,{\bf
n}\,\right)$ as seen by the Eulerian observers comoving with ${\bf
n}$:
\be
\l{zeromag}
H_{ab}({\bf n}) = 0 \ .
\ee
This has the important implication: there can exist {\em no\/}
gravitational radiation in a quasi-Newtonian cosmology
\ct{vanetal97}. Being an invariant physical feature, it must also
hold with respect to all other observers {\em different\/} from the
Eulerian ones comoving with ${\bf n}$. This shows that, considered
as General Relativity spacetimes, such cosmological models are very
special, for in fact generic motion of matter {\em will\/} generate
gravitational radiation. However, its absence, of course,
corresponds to the situation encountered in the Newtonian theory of
gravitation.

\subsubsection{Acceleration scalar potential}
Preservation of the irrotationality condition (\r{zerovor})
requires from the vorticity evolution equation
(\,cf. Refs. \ct{ell71} and \ct{hve96}\,) that $0 =
\epsilon^{abc}\,\mbox{D}_b\dot{n}_c$;\footnote{In the {\em
comoving\/} irrotational case of a barotropic perfect fluid, where
the fluid acceleration is proportional to the spatial gradient of
the total energy density, this is an identity. Here, however, it
provides a {\em genuine\/} constraint on $\dot{n}^{a}$.} the
spatial rotation of the relativistic acceleration $\dot{n}^{a}({\bf
n}) := n^{b}\nabla_{b}n^{a}$ of the timelike reference congruence
${\bf n}$ must vanish. Consequently, the acceleration is
proportional to the spatial gradient of a $1+3$ covariantly defined
scalar field $N$ \ct{yor79}:
\be
\l{acc}
\ndot_{a}({\bf n}) := N^{-1}\,\D_{a}N \ .
\ee
As already mentioned, this relativistic acceleration scalar
potential corresponds to the Newtonian gravitational scalar
potential. For convenience one may alternatively introduce a
dimensionless scalar potential $\Phi :=\ln (N/N_{0})$, where
$N_{0}$ is a constant.\footnote{The physical dimension of the lapse
function $N$ is $\lgth$.}

Taken together, the physical properties expressed by
Eqs. (\r{zerovor}) - (\r{acc}) clearly justify the terminology
`quasi-Newtonian' for the cosmological models at hand.\footnote{For
further discussion see Ref. \ct{bar80}, and the discussion of
`Newtonian-like' spacetimes in Ref. \ct{maaetal98}.}

\subsection{Local coordinates}
The relativistic acceleration scalar potential $N$ is nothing but
the prominent lapse function of the ADM $3+1$ formalism, which
embodies (part of) one's personal choice of how to `move forward in
time' \ct{mtw73} within a dynamical spacetime setting. In terms of
a set of (dimensionless) local coordinates $\{\,x^{i}\,\}$, the
infinitesimal line element of a quasi-Newtonian cosmology can be
written in the form (\,see, e.g., Ref. \ct{treell71}\,)\footnote{In
the case when $S = S(t)$, we can alternatively use the conformal
time coordinate $\eta(t) = \int_{t_{0}}^{t} dy/S(y)$ of physical
dimension $\lgth^{-1}$, integrated along the integral curves of the
timelike reference congruence ${\bf n}$, and the line element takes
the form $ds^{2} = S^{2}(\eta)(\,-\,N^{2}(x^{i})\,d\eta^{2} +
d\sig^{2}\,)$. Note, however, this does not work in general when $S
= S(t,x^{\alpha})$; for then $\eta = \eta(t,x^{\alpha})$ and there
will be time--space cross terms in the resultant form of the line
element.}
\be
\l{qnds2}
ds^{2} = -\,N^{2}(x^{i})\,dt^{2} + S^{2}(x^{i})\,d\sig^{2} \ ,
\hsp5 d\sig^{2} := f_{\alpha\beta}(x^{\gam})\,dx^{\alpha}\,
dx^{\beta} \ ,
\hsp5 n^{a} = N^{-1}\,\d^{a}\!_{0} \ .
\ee
(the time coordinate is $t = x^{0}$; spacetime coordinate indices
$i, j, k, \dots$ run through 0 -- 3, and spatial coordinate indices
$\alpha, \beta, \gam, \dots$ run through 1 -- 3), where the rate of
expansion scalar $\Th$ of the normals ${\bf n}$ is given by
\be
\l{exp}
\Th(x^{i}) = 3\,\frac{\dot{S}}{S}
= 3\,(SN)^{-1}\,\frac{\partial S}{\partial t} \ ,
\ee
and $S$, as usual, denotes a representative expansion length
scale. Because this form of the line element implies that ${\bf n}$
is irrotational and shearfree, {\em existence of local coordinates
such that the line element has the form (\r{qnds2}) is a necessary
and sufficient condition that a cosmology is quasi-Newtonian in our
sense\/} (note that this form includes both the FLRW and
Schwarzschild spacetimes).

\subsection{The matter flow}
The dust matter flow of our cosmological model\footnote{In our
discussion, for simplicity, we set the cosmological term to zero,
$\Lambda = 0$.} is moving with average 4-velocity ${\bf \tilde{u}}$
($\tilde{u}_{a}\,\tilde{u}^{a} = -\,1$). Thus, relative to the
Lagrangean reference frame comoving with ${\bf \tl{u}}$ (\,where
$\dot{\tl{u}}{}^{a}({\bf \tl{u}}) :=
\tl{u}{}^{b}\nabla_{b}\tl{u}{}^{a} = 0$\,), the
energy-momentum-stress tensor takes the familiar form
\be
\l{emom1}
T_{ab}({\bf \tl{u}}) = \tl{\mu}\,\tl{u}_{a}\,\tl{u}_{b} \ .
\ee
$1+3$-decomposing ${\bf \tl{u}}$ with respect to the Eulerian ${\bf
n}$-frame we obtain \ct{hveugg97,bruetal92}
\be
\l{tlu}
\tilde{u}^{a} := \gam\,(n^{a}+v^{a}) \ , \hsp5
\gam:=(1-v^{2})^{-1/2} \ , \hsp5 v^{2}:=v_{a}\,v^{a} \geq 0 \ ,
\hsp5 v_{a}\,n^{a} = 0 \ . 
\ee
The variable $v^{a}$ is the {\em peculiar velocity\/} of the matter
flow with respect to the ${\cal T}$:
$\left\{t=\mbox{const}\right\}$. Hence, the $1+3$ covariant
decomposition relative to the Eulerian ${\bf n}$-frame (\,where
$U^{a}{}_{b}:=-\,n^{a}\,n_{b} $,
$h^{a}{}_{b}=\delta^{a}{}_{b}-U^{a}{}_{b}$\,) leads to
\be
\l{emom2}
T_{ab}({\bf n}) = \mu\,n_{a}\,n_{b} + 2\,q_{(a}\,n_{b)}
+ p\,h_{ab} + \pi_{ab} \ ;
\ee
the matter variables are given by \ct{hveugg97}\footnote{We could
consider several matter components: each would obey similar
equations, with the total matter variables simply being the sum of
those for each component. Note the direct relation to the approach
taken in formulating thermodynamical properties of matter within
kinetic theory.}
\be
\l{mat}
\mu  =  \gam^{2}\,\tilde{\mu} \ , \hsp5
p = \sfrac{1}{3}\,\mu\,v^{2} \ , \hsp5
q^{a}  = \mu\,v^{a} \ , \hsp5
\pi_{ab} = \mu\,v_{\lgl a}\,v_{b\rgl} \ .
\ee
Note that $p$ and $\pi_{ab}$, which are typically non-Newtonian
sources of spacetime curvature, are {\em quadratic\/} in $v^{a}$,
so they will be negligible when we linearise our models about a
non-comoving FLRW background, assuming that $v^{a}$ itself is a
deviation variable of first-order smallness. However, $q^{a}$ is
${\em linear\/}$ in $v^{a}$; this will be crucial later on. The
dependence of the fluid kinematical variables $\tl{\sig}_{ab}$,
$\tl{\Th}$ and $\tl{\om}_{ab}$ on the set $\{\,n^{a}, \,\ndot^{a},
\,\Th, \,v^{a}, \,\D_{a}v_{b}\,\}$ is given explicitly in
subsection \r{subsec:matuder} of the appendix.\footnote{For the
covariant derivatives of $n^{a}$ and $v^{a}$ one obtains,
respectively, $\nabla_{a}n_{b} = -\,n_{a}\,\dot{n}_{b} +
\sfrac{1}{3}\,\Th\,h_{ab}$ and $\nabla_{a}v_{b} =
-\,(v_{c}\dot{n}^{c})\,n_{a}\,n_{b} - n_{a}\,\dot{v}_{\langle
b\rangle } + \sfrac{1}{3}\,\Th\,v_{a}\,n_{b} + \D_{a}v_{b}$.} In
the limit $v^{a} \rightarrow 0$ the situation at hand reduces to
the standard comoving description of FLRW cosmological models with
dust matter source field (in this limit, the acceleration of the
normals has to vanish, because the dust flow lines are geodesic).

\subsection{Uniqueness of $N$ and ${\bf n}$}
\l{subsec:qnunique}
As is well-known, the ADM $3+1$ formalism does {\em not\/} provide
one with a natural evolution equation for the lapse function $N$
(nor the shift vector $N^{a}$), and, therefore, by Eq. (\r{acc}),
also {\em not\/} for the acceleration $\dot{n}^a$ of the normal
congruence ${\bf n}$ \ct{adm62,yor79}. Hence, it is a major
dynamical issue as to how to determine, for a given physical
problem, the best prescription for moving forward in coordinate
time $t$ from one spacelike 3-surface within the foliation to the
next \ct{yor79}. In general, a `time slicing condition' leading to
a typically differential constraint relation has to be introduced
by hand.\footnote{This indeterminacy does not arise when the
congruence of preferred worldlines is comoving with well-specified
matter (e.g., a perfect fluid with specified equation of
state). However, that is not the case here.} However, a fundamental
point has already been raised above: despite this lack of
uniqueness of $N$ from a dynamical viewpoint, we expect the normal
congruence ${\bf n}$ in our case to be highly restricted {\em
geometrically\/}, indeed to usually be uniquely determined, because
of the conditions (a) that Eulerian observers comoving with ${\bf
n}$ will see zero magnetic Weyl curvature, Eq. (\r{zeromag}), and
(b) have a highly restricted electric Weyl curvature, or,
equivalently, a very restricted intrinsic curvature allowed for the
spacelike 3-surfaces in the family ${\cal T}$:
$\left\{t=\mbox{const}\right\} $ (this will be explored below). The
hope then is that choice of this congruence essentially determines
the time evolution of $N$, because it determines the acceleration
$\dot{n}^{a}$ (\,which occurs in Eq. (\r{acc})\,); so uniqueness of
${\bf n}$ will imply a unique choice of the lapse function.

Consequently, it is important to determine when the Newtonian-like
timelike reference congruence ${\bf n}$ really {\em is\/}
unique. As regards point (a) above, if we express the Weyl
curvature relative to ${\bf n}$ in terms of its electric part, we
obtain
\be
\l{weyln}
C^{ab}{}_{cd}({\bf n}) = 4\,n^{[a}\,n_{[c}\,E^{b]}{}_{d]}
+ 4\,h^{[a}{}_{[c}\,E^{b]}{}_{d]} \ .
\ee
Hence, the magnetic Weyl curvature relative to {\em another\/}
timelike congruence, ${\bf\tl{n}}$, where $\tl{n}^{a} =
\gam\,(n^{a}+v^{a})$ (\,cf. Eq. (\r{tlu})\,), is
\be
\l{magtln}
\tl{H}_{ab}({\bf\tl{n}}) = 2\,\gam^{2}\,\eps_{cd\lgl a}\,
[\ n_{b\rgl}\,(E^{c}\!_{e}v^{e}) + E^{c}\!_{b\rgl}\ ]\,v^{d} \ .
\ee
So, generically, for $E_{ab} \neq 0$ and $v^{a} \neq 0$, we have
$\tl{H}_{ab} \neq 0$. On the other hand, both $H_{ab}({\bf n}) = 0$
and $\tl{H}_{ab}({\bf\tl{n}}) = 0$, iff
\be
\l{zerotlmag}
\eps_{cd\lgl a}\,[\ n_{b\rgl}\,(E^{c}\!_{e}v^{e})
+ E^{c}\!_{b\rgl}\ ]\,v^{d} = 0 \ .
\ee
This will be satisfied in the FLRW case, where $0 = E_{ab}({\bf n})
= H_{ab}({\bf n})$ (implying $0 = \tilde{E}_{ab}({\bf \tilde{n}}) =
\tilde{H}_{ab}({\bf \tilde{n}})$; cf. subsection \r{subsec:matuder}
of the appendix), and also for cosmologies with even shearing
timelike reference congruences that possess LRS symmetry, if
$v^{a}$ characterises a Lorentz-boost along the $1+3$ covariantly
defined preferred spacelike direction ${\bf e}$
\ct{ell67,hveell96}. Equation (\r{zerotlmag}) is a necessary (but
not sufficient) condition that the same spacetime geometry can
appear Newtonian-like relative to {\em two\/} (or more) timelike
congruences. When this is satisfied, one then also needs to find if
(b) is satisfied: does the electric Weyl curvature have the highly
restricted form that allows a zero initial rate of shear to remain
zero, given the matter source fields present? These conditions are
rather complex, nevertheless, we would like to get a complete
solution:
\begin{quotation}
{\bf Problem 1}: Determine all quasi-Newtonian cosmologies that
allow more than one Newtonian-like timelike congruence.
\end{quotation}
This family includes constant curvature spacetimes and the FLRW
spacetimes; are there any others?

\section{$1+3$ covariant dynamical equations}
\l{sec:13covqn}
Insisting on the timelike reference congruence ${\bf n}$ to have
zero rate of shear, $\sig_{ab}=0$, converts the original
$\dot{\sig}^{\langle ab\rangle}$-equation (a Ricci identity
determining the evolution of $\sig_{ab}$ along ${\bf n}$) into a
constraint equation which we can regard as {\em either\/}
restricting the allowed time slicing, given the Weyl curvature of
$\left(\,{\cal M},\,{\bf g},\,{\bf n}\,\right)$ (i.e., as a
differential time slicing condition for the lapse function $N$),
{\em or\/} as restricting the allowed electric Weyl curvature,
$E_{ab}({\bf n})$, given a choice of the time slicing of
$\left(\,{\cal M},\,{\bf g},\,{\bf n}\,\right)$. In either case,
the constraint so generated assumes the explicit form\footnote{We
draw the attention of the reader to the fact that the differential
time slicing condition constituted by Eq. (\r{slicing}) is
conceptually analogous to obtaining a relativistic generalisation
of the PFE when imposing `maximal' time slicing characterised by
$\Th =0 \Rightarrow \dot{\Th} = 0$ (but $\sig_{ab} \neq 0$) in the
case of asymptotically flat spacetimes \ct{yor79}.}
\be
\l{slicing}
0 = E^{ab} - \D^{\lgl a}\ndot^{b\rgl} - \ndot^{\lgl a}\,
\ndot^{b\rgl} - \sfrac{1}{2}\,\mu\,v^{\lgl a}\,v^{b\rgl} \ .
\ee
This is a highly restrictive assumption; we wish to explore its
implications.

Using the variables specified in Eq. (\r{mat}), under the stated
assumptions (\r{zerovor}) - (\r{acc}) characterising
quasi-Newtonian cosmologies with dust matter source field one
derives from the general $1+3$ covariant dynamical equations
(\,cf. Refs. \ct{ell71} and \ct{hve96}\,) the set: \enl

\noindent
{\em Evolution equations\/}:
\bea
\l{covthdot}
\dot{\Th} & = & -\,\sfrac{1}{3}\,\Th^{2} + \D_{a}\ndot^{a}
+ (\ndot_{a}\ndot^{a}) - \sfrac{1}{2}\,(1+v^{2})\,\mu \\
\l{covmudot}
\dot{\mu} & = & -\,(1+\sfrac{1}{3}v^{2})\,\Th\,\mu
- v^{a}\D_{a}\mu - \mu\,\D_{a}v^{a} - 2\,\mu\,(v_{a}\ndot^{a}) \\
\l{covvdot}
\dot{v}{}^{\lgl a\rgl} & = & -\,\sfrac{1}{3}\,(1-v^{2})\,\Th\,v^{a}
- v^{b}\D_{b}v^{a} + (v_{b}\ndot^{b})\,v^{a}
- \ndot^{a} \\
\l{covedot}
\dot{E}{}^{\lgl ab\rgl} & = & -\,\Th\,E^{ab}
+ \sfrac{2}{3}\,(1-\sfrac{1}{4}v^{2})\,\Th\,\mu\,v^{\lgl a}
\,v^{b\rgl} + \sfrac{1}{2}\,(v^{c}\D_{c}\mu)\,v^{\lgl a}\,v^{b\rgl}
- \sfrac{1}{2}\,v^{\lgl a}\,\D^{b\rgl}\mu \nonumber \\
& & \hsp5 + \ \sfrac{5}{6}\,(\D_{c}v^{c})\,\mu\,
v^{\lgl a}\,v^{b\rgl}
- \sfrac{1}{2}\,\mu\,\D^{\lgl a}v^{b\rgl}
+ \mu\,v_{c}\,v^{\lgl a}\,\D^{\lgl b\rgl}v^{c\rgl}
+ \mu\,v_{c}\,v^{\lgl a}\,\D^{[b\rgl}v^{c]} \ .
\eea
The evolution equation for $v^{a}$ is a consequence of the original
$\dot{q}^{\langle a\rangle}$-equation (Bianchi identity); it
establishes conservation of momentum (density). As such it is the
relativistic analogue of the Euler equation in the Newtonian
theory. \enl

\noindent
{\em Primary constraint equations\/}:
\bea
\l{covdivsig}
0 & = & (C_{1})^{a} \ := \ \D^{a}\Th - \sfrac{3}{2}\,\mu\,v^{a} \\
\l{covomdot}
0 & = & (C_{2})^{a} \ := \ \eps^{abc}\,\D_{b}\ndot_{c} \\
\l{covsigdot}
0 & = & (C_{3})^{ab} \ := \ \ E^{ab} - \D^{\lgl a}\ndot^{b\rgl}
- \ndot^{\lgl a}\,\ndot^{b\rgl}
- \sfrac{1}{2}\,\mu\,v^{\lgl a}\,v^{b\rgl} \\
\l{covdive}
0 & = & (C_{4})^{a} \ := \ \D_{b}E^{ab}
- \sfrac{1}{3}\,(1+\sfrac{1}{2}v^{2})\,\D^{a}\mu
+ \sfrac{1}{3}\,[\ \Th + \sfrac{3}{2}\,\mu^{-1}\,(v^{b}\D_{b}\mu)
+ \sfrac{5}{3}\,(\D_{b}v^{b})\ ]\,\mu\,v^{a} \nonumber \\
& & \hspace{2cm} + \ \sfrac{1}{6}\,\mu\,v_{b}\,
[\ \D^{\lgl a}v^{b\rgl} - 5\,\D^{[a}v^{b]}\ ] \ .
\eea
The constraint $(C_{1})^{a}$, being the reduced form of the
original $\D_{b}\sig^{ab}$-constraint (Ricci identity), states that
the peculiar motion of the matter as quantified by $v^{a}$ induces
a non-zero spatial gradient in the expansion rate $\Th$ of the
normals ${\bf n}$. Constraints $(C_{2})^{a}$ and $(C_{3})^{ab}$ are
the key new constraints resulting from the quasi-Newtonian
assumptions, $(C_{2})^{a}$ leading to the existence of the
relativistic acceleration scalar potential $N$ (\,see
Eq. (\r{acc})\,). $(C_{4})^{a}$ is the $\D_{b}E^{ab}$-constraint
(Bianchi identity). Additionally, there are constraints related to
$H_{ab}$:
\bea
\l{covhcons}
0 & = & (C_{H})^{ab} \ := \ H^{ab} \\
\l{covdivh}
0 & = & (C_{D\cdot H})^{a} \ := \ \D_{b}(C_{H})^{ab}
+ \sfrac{1}{2}\,\eps^{abc}\,\D_{b}(\mu\,v_{c}) \ .
\eea
$(C_{H})^{ab}$ holds as long as Eqs. (\r{zerovor}) and
(\r{zerosig}) apply. Using Eq. (\r{zerocom}) of subsection
\r{subsec:com} in the appendix, the $(C_{D\cdot H})^{a}$-constraint
(Bianchi identity) can be reduced to $0 = \D_{b}(C_{H})^{ab} -
\sfrac{1}{3}\,\eps^{abc}\,\D_{b}(C_{1})_{c}$ and thus is
identically satisfied if each of $(C_{1})^{a}$ and $(C_{H})^{ab}$
is valid. It then follows that
\be
\l{pvrot}
\eps^{abc}\,\D_{b}v_{c} = \eps^{abc}\,v_{b}\,\frac{\D_{c}\mu}{\mu}
\hsp5 \Leftrightarrow \hsp5
\D_{[a}v_{b]} = \mu^{-1}\,v_{[a}\,\D_{b]}\mu \ ;
\ee
the spatial rotation of $v^{a}$ is equal to the algebraic cross
product between $v^{a}$ itself and the fractional total energy
density gradient $(\D_{a}\mu)/\mu$. As such, the vorticity of
$v^{a}$ will be of {\em second\/}-order smallness when we only
consider first-order deviations about a FLRW background setting
associated with the ${\cal T}$: $\left\{t=\mbox{const}\right\}$.
Altogether, the fully orthogonally projected covariant derivative
of $v^{a}$ can now be decomposed as
\be
\l{pvdec}
\D_{a}v_{b} = \D_{\lgl a}v_{b\rgl} + \sfrac{1}{3}\,(\D_{c}v^{c})\,
h_{ab} + \mu^{-1}\,v_{[a}\,\D_{b]}\mu \ ;
\ee
$\D_{\lgl a}v_{b\rgl}$ and $\D_{a}v^{a}$ representing the
distortion (or shear) and spatial divergence (or expansion) of the
peculiar velocity field, respectively. With condition
(\r{covhcons}) prescribed, the status of the $\dot{H}^{\lgl
ab\rgl}$-equation (Bianchi identity) can take either the rank of
yet another constraint relation (\,cf. Ref. \ct{vanetal97}\,), or
it naturally yields the propagation of $(C_{H})^{ab}$ along ${\bf
n}$. Here we find that
\be
\l{chdot}
(C_{\dot{H}})^{ab} = (\dot{C}_{H})^{\lgl ab\rgl}
= -\,\Th\,(C_{H})^{ab} - [\ \D^{\lgl a} + \ndot^{\lgl a}\ ]\,
(C_{2})^{b\rgl} - \eps^{cd\lgl a}\,[\ \D_{c} + \ndot_{c}\ ]\,
(C_{3})^{b\rgl}\!_{d}
\ ;
\ee
if $(C_{2})^{a}$ and $(C_{3})^{ab}$ are satisfied on ${\cal
T}_{0}$: $\left\{t_{0}=\mbox{const}\right\}$, then a consistent set
of initial data for quasi-Newtonian cosmologies with dust matter
source field characterised by Eqs. (\r{zerovor}) - (\r{acc})
generates {\em no\/} magnetic Weyl curvature along ${\bf n}$.

A particularly important restriction on $\left(\,{\cal M},\,{\bf
g},\,{\bf n}\,\right)$ results as a compatibility requirement
between $(C_{3})^{ab}$ and $(C_{4})^{a}$. It follows that
\be
0 = \D_{b}(C_{3})^{ab} = (C_{4})^{a} - \ndot_{b}\,(C_{3})^{ab}
- \sfrac{2}{3}\,(C_{5})^{a} \ ,
\ee
where $(C_{5})^{a}$ is given by
\bea
\l{covc5}
(C_{5})^{a} & := & \ \D^{a}\,[\ \D_{b}\ndot^{b}
+ (\ndot_{b}\ndot^{b})\ ] + \ndot^{a}\,[\ \D_{b}\ndot^{b}
+ (\ndot_{b}\ndot^{b}) - \sfrac{1}{3}\,\Th^{2}
+ (1-\sfrac{1}{2}v^{2})\,\mu\ ] \nonumber \\
& & \hsp5 - \ \sfrac{1}{2}\,(1+v^{2})\,\D^{a}\mu
+ \sfrac{1}{2}\,[\ \Th + 3\,\mu^{-1}\,(v^{b}\D_{b}\mu)
+ \sfrac{10}{3}\,(\D_{b}v^{b})
+ 3\,(v_{b}\ndot^{b})\ ]\,\mu\,v^{a} \nonumber \\
& & \hsp5 + \ \sfrac{1}{2}\,\mu\,v_{b}\,
[\ \D^{\lgl a}v^{b\rgl} - 5\,\D^{[a}v^{b]}\ ] \ ,
\eea
and $0 = (C_{5})^{a}$ is demanded. Using Eq. (\r{acc}), this
provides a third-order partial differential equation for $N$.

In summary, the particular non-trivial constraints characterising
the quasi-Newtonian case at hand are $(C_{3})^{ab}$ and its
immediate consequence $(C_{5})^{a}$, for these embody the
consequences of the zero-shear requirement (\r{zerosig}).

\subsection{The geometry of the orthogonal spacelike 3-surfaces}
Next, we give the geometrical variables characterising the
intrinsic curvature properties of the orthogonal spacelike
3-surfaces ${\cal T}$: $\left\{t=\mbox{const}\right\}$, which
exist because of the requirement (\r{zerovor}). \enl

\noindent
{\em Trace and tracefree parts of Gau\ss\ embedding equation\/}:
\bea
\l{covfried}
0 & = & (C_{G}) \ := \ {}^{3}\!R
+ \sfrac{2}{3}\,\Th^{2} - 2\,\mu \\
\l{covan3ric}
0 & = & (C_{G})^{ab} \ := \ {}^{3}\!S^{ab} - E^{ab}
- \sfrac{1}{2}\,\mu\,v^{\lgl a}\,v^{b\rgl} \ .
\eea
${}^{3}\!R$ and ${}^{3}\!S_{ab}$ denote the respective irreducible
parts of the 3-Ricci curvature. As demonstrated for the example of
a Lagrangean description of an irrotational dust matter source
field in Ref. \ct{vel97}, the Gau\ss\ embedding equation has to be
included in order to obtain a set of ($1+3$ covariant) dynamical
equations which is equivalent to the EFE. Note that the geometrical
structure of Eq. (\r{covfried}) is identical to the Friedmann
equation in FLRW cosmology.\enl

\noindent
{\em 3-Cotton--York tensor\/} \ct{hveugg97}:
\bea
\l{covan3cy}
{}^{3}\!C^{ab} & := & h^{1/3}\,\eps^{cd\lgl a}\,\D_{c}
{}^{3}\!S^{b\rgl}\!_{d} \nonumber \\
& = & -\,2\,h^{1/3}\,\eps^{cd\lgl a}\ [\ \ndot_{c}\,
\D^{\lgl b\rgl}\ndot_{d\rgl} + \sfrac{1}{2}\,\mu\,v_{c}\,
(\D^{\lgl b\rgl}v_{d\rgl} - \D^{[b\rgl}v_{d]})\ ]
- h^{1/3}\,\mu\,v^{\lgl a}\,\eps^{b\rgl cd}\,\ndot_{c}\,v_{d}
\nonumber \\
& & \hsp5 + \ h^{1/3}\,\eps^{cd\lgl a}
\ [\ \D_{c}(C_{G})^{b\rgl}\!_{d} + \D_{c}(C_{3})^{b\rgl}\!_{d}
- \ndot_{c}\,(C_{3})^{b\rgl}\!_{d}\ ] + h^{1/3}\,\ndot^{\lgl a}\,
(C_{2})^{b\rgl} \ .
\eea
The 3-Cotton--York tensor, although {\em not\/} having the
dimension of a curvature variable, encodes the conformal curvature
properties of the ${\cal T}$: $\left\{t=\mbox{const}\right\}$.
They are conformally flat if ${}^{3}\!C_{ab}$ is zero. It has been
suggested that a non-zero 3-Cotton--York tensor could be intimately
related to the presence of gravitational radiation in a spacetime
geometry (\,see, e.g., Ref. \ct{wai79}\,). Consequently, one would
expect it to vanish in the quasi-Newtonian case.

\subsection{Propagation of $N$ and $\ndot^{a}$}
Having now attained the full set of $1+3$ covariant dynamical
equations for quasi-Newtonian cosmlogies with dust matter source
field, it is clear (in accordance with the ADM $3+1$ results) that
there is no explicit equation determining the evolution of $N$ or
$\ndot^{a}$. Two differential expressions containing the latter
particularly spring to the eye. In terms of $N$ the
$\ndot^{a}$-part of Eq. (\r{slicing}) is given by
\be
\l{accdis}
\D_{\lgl a}\ndot_{b\rgl} + \ndot_{\lgl a}\,
\ndot_{b\rgl} = N^{-1}\,\D_{\lgl a}\D_{b\rgl}N \ .
\ee
The symmetric tracefree derivative $\D_{\lgl a}\ndot_{b\rgl}$ is
the distortion of $\ndot^{a}$. On the other hand, the spatial
divergence $\D_{a}\ndot^{a}$ can be interpreted as its
expansion. Again, in terms of $N$ we have
\be
\l{nlapl}
\D_{a}\ndot^{a} + (\ndot_{a}\ndot^{a}) = N^{-1}\,\D_{a}\D^{a}N \ ;
\ee
this is a source term in Eq. (\r{covthdot}) and corresponds to the
Laplacian of the gravitational scalar potential in the Newtonian
theory. We also have no evolution equations for neither of these
two differential quantities. This remains true if one writes out
the full set of equations in an associated $1+3$ orthonormal frame
formalism, which we give in subsection \r{subsec:13onfqn} of the
appendix. Hence, we face:
\begin{quotation}
{\bf Problem 2}: In those cases where the Newtonian-like timelike
congruence is unique, how is the corresponding evolution of the
lapse function $N$ along these worldlines determined?
\end{quotation}
Lacking explicit evolution equations for any of the relativistic
acceleration,\footnote{This contrasts the case of a barotropic
perfect fluid represented in a {\em comoving\/} frame, where (i)
the relativistic acceleration is determined from the spatial
gradient of the pressure by the momentum conservation equation,
(ii) the pressure is determined from the total energy density via
the equation of state, and (iii) the energy density time evolution
is determined by the energy conservation equation. In the present
situation, because the reference frame is {\em not\/} comoving,
step (i) fails; given any specified acceleration distribution, the
momentum conservation equation determines the peculiar velocity of
the matter fluid relative to the Eulerian reference frame.} its
spatial divergence,\footnote{In the {\em strictly\/} Newtonian case
we would additionally have $\Th = 0$, and then the Raychaudhuri
equation (\r{covthdot}) becomes the relativistic generalisation of
the PFE determining this spatial divergence. In the case of an
expanding cosmological model, however, this equation rather
determines the time rate of change of the expansion that is
compatible with whatever value holds for this spatial divergence.}
or other spatial derivatives, we presume the former must be
implicit in the consistency conditions that arise when we impose
the quasi-Newtonian restrictions (\r{zerovor}) and (\r{zerosig}) on
our timelike reference congruence. However, we have been unable to
obtain a general prescription as to how this happens.

\subsubsection{An Ansatz}
In a formal sense, an evolution equation for $\ndot^{a}$ in the
quasi-Newtonian dust case can be derived by combining Eq. (\r{acc})
with the commutator relation (\r{3sclcom}) in subsection
\r{subsec:com} of the appendix, thus yielding
\be
\l{accdot}
(\ndot)\,{}\dot{}\,{}^{\lgl a\rgl} = N^{-1}\,\D^{a}(\dot{N})
- \sfrac{1}{3}\,\Th\,\ndot^{a} \ .
\ee
The latter makes explicit the necessity for an evolution equation
for $N$, required to close the set of dynamical equations. Failing
any other way of determining this, we propose to evolve $N$ along
${\bf n}$ according to the {\em Ansatz\/}
\be
\l{ndot}
\dot{N} = \alpha\,\Th\,N \hsp5 \mbox{where} \hsp5
\alpha = \mbox{const} \ ,
\ee
which is a common choice when modelling dynamical spacetime
geometries by numerical methods \ct{bonetal95}; for $\alpha = 1$,
e.g., this Ansatz specialises to the so-called `harmonic gauge'
where $N$ propagates along ${\bf n}$ via a wave equation at the
speed of light (\,cf. Refs. \ct{abretal97} and
\ct{bonetal95}\,). Other particular values for the parameter
$\alpha$ will be mentioned in section \r{sec:qnflrwlin} below. With
this assumption, it follows that
\be
\l{covaccdot}
(\ndot)\,{}\dot{}\,{}^{\lgl a\rgl} = (\alpha-\sfrac{1}{3})\,
\Th\,\ndot^{a} + \alpha\,\sfrac{3}{2}\,\mu\,v^{a}
+ \alpha\,(C_{1})^{a} \ .
\ee
Thus, the evolution of the dynamical quasi-Newtonian cosmologies is
now deterministic.

\section{Constraint analysis}
\l{sec:qncons}
To check whether quasi-Newtonian cosmologies with dust matter
source field can provide consistent solutions of the EFE requires a
detailed analysis of the preservation properties of the constraint
relations as a model evolves. That all constraints involved be
preserved along the integral curves of ${\bf n}$ is a {\em
necessary\/} condition for consistency of the $1+3$ covariant
dynamical equations; a {\em full\/} consistency check involves
additionally showing, e.g., that all Jacobi and Ricci equations of
an associated $1+3$ orthonormal frame formalism are satisfied
(\,see, e.g., Refs. \ct{ell67} and \ct{hveugg97}\,), or casting the
evolution equations into a first-order symmetric hyperbolic form
(\,see, e.g., Refs. \ct{abretal97} and \ct{fri98} for the cases of
a vacuum and a perfect fluid spacetime geometry, respectively\,).

As mentioned, ideally we would like to derive the form of $\dot{N}$
in the generic case from this consistency analysis, but have been
unable to do so; consequently, a prescription for $\dot{N}$ is
needed, followed by a subsequent consistency check. We assume the
evolutionary behaviour of the lapse function $N$ along the integral
curves of ${\bf n}$ to be given by the Ansatz (\r{ndot}). Using
relations for the propagation of spatial derivative terms given in
Ref. \ct{hve96}\footnote{Note the error in Eq. (B.13) of
Ref. \ct{hve96}: the numerical factor preceding the last term on
the RHS should be `$3$' rather than `$2$'.} and also
Eq. (\r{zerocom}) in subsection \r{subsec:com} of the appendix, we
then find that the primary constraints (\r{covdivsig}) -
(\r{covdive}) and (\r{covc5}) propagate according to
\bea
\l{c1dot}
(\dot{C}_{1})^{\lgl a\rgl} & = & - \,\Th\,(C_{1})^{a}
+ (C_{5})^{a} \\
\l{c2dot}
(\dot{C}_{2})^{\lgl a\rgl} & = & (\alpha-\sfrac{2}{3})\,\Th
\,(C_{2})^{a}
+ \sfrac{1}{3}\,\eps^{abc}\,\ndot_{b}\,(C_{1})_{c} \\
\l{c3dot}
(\dot{C}_{3})^{\lgl ab\rgl} & = & -\,\Th\,(C_{3})^{ab}
- \alpha\,\D^{\lgl a}(C_{1})^{b\rgl} - 4\,(\alpha-\sfrac{1}{12})\,
\ndot^{\lgl a}(C_{1})^{b\rgl} \nonumber \\
& & \hsp5 - \ (\alpha+\sfrac{1}{3})\,
[\ \Th\,\D^{\lgl a}\ndot^{b\rgl} + \sfrac{3}{2}\,\mu\,\D^{\lgl
a}v^{b\rgl} + \sfrac{3}{2}\,v^{\lgl a}\,\D^{b\rgl}\mu\ ]
- 3\,(\alpha-\sfrac{1}{9})\,\Th\,\ndot^{\lgl a}\,\ndot^{b\rgl}
\nonumber \\
& & \hsp5 - \ 6\,(\alpha-\sfrac{1}{3})\,\mu\,\ndot^{\lgl a}\,
v^{b\rgl} + 2\,\mu\,v_{c}\,v^{\lgl a}\,
[\ \D^{\lgl b\rgl}v^{c\rgl} + \D^{[b\rgl}v^{c]}\ ] \nonumber \\
& & \hsp5 + \ [\ (1-\sfrac{1}{3}v^{2})\,\Th
+ \mu^{-1}\,(v^{c}\D_{c}\mu) + \sfrac{5}{3}\,(\D_{c}v^{c})\ ]\,
\mu\,v^{\lgl a}\,v^{b\rgl} \\
\l{c4dot}
(\dot{C}_{4})^{\lgl a\rgl} & = & - \,\sfrac{4}{3}\,\Th\,(C_{4})^{a}
+ \sfrac{1}{3}\,(1+v^{2})\,\mu\,(C_{1})^{a} - \sfrac{1}{6}\,
\mu\,v^{a}\,[\,v_{b}\,(C_{1})^{b}\,] - E^{a}\!_{b}\,(C_{1})^{b}
+ \sfrac{1}{3}\,\mu\,(C_{6})^{a} \\
\l{c5dot}
(\dot{C}_{5})^{\lgl a\rgl} & = & (\alpha-1)\,\Th\,(C_{5})^{a}
+ \alpha\,[\ \D^{a} + 2\,\ndot^{a}\ ]\,[\,\D_{b}(C_{1})^{b}\,]
+ 4\,(\alpha-\sfrac{1}{12})\,[\ \D^{a} + 2\,\ndot^{a}\ ]\,
[\,\ndot_{b}(C_{1})^{b}\,] \nonumber \\
& & \hsp5 + \ 2\,[\ (\alpha-\sfrac{1}{3})\,
\D_{b}\ndot^{b} + 2\,(\alpha-\sfrac{1}{12})\,(\ndot_{b}\ndot^{b})
- \alpha\,\sfrac{1}{6}\,\Th^{2} \nonumber \\
& & \hspace{2cm} + \ \sfrac{1}{2}\,(\alpha
+ \sfrac{1}{2}(\alpha-2)v^{2} - \sfrac{1}{6}v^{4})\,\mu\ ]\,
(C_{1})^{a} \nonumber \\
& & \hsp5 + \ \sfrac{3}{2}\,(\alpha-\sfrac{5}{3}+\sfrac{1}{3}v^{2})
\,\mu\,v^{a}\,[\,v_{b}(C_{1})^{b}\,] + \sfrac{3}{2}\,(\alpha
+\sfrac{1}{3}+\sfrac{5}{9}v^{2})\,\mu\,\D^{a}(\D_{b}v^{b})
\nonumber \\
& & \hsp5 + \ \sfrac{1}{2}\,(\alpha + \sfrac{1}{3} + (\alpha
+\sfrac{4}{3})v^{2} - \sfrac{1}{3}v^{4})\,\Th\,\D^{a}\mu
- (\alpha+\sfrac{1}{3})\,\Th^{2}\,\mu\,v^{a} \nonumber \\
& & \hsp5 + \ \sfrac{3}{2}\,(\alpha+\sfrac{1}{3}
+ (\alpha-\sfrac{16}{9})v^{2} + \sfrac{8}{9}v^{4})\,\mu^{2}\,v^{a}
+ \mbox{further non-linear terms} \ ,
\eea
where $(C_{6})^{a}$ is defined by
\bea
\l{covc6}
(C_{6})^{a} & := & \D^{a}(\D_{b}v^{b}) - \sfrac{3}{2}\,
\D_{b}\D^{\lgl a}v^{b\rgl} \nonumber \\
& & \hsp5 - \ [\ \sfrac{1}{3}\,\Th^{2} - (1-\sfrac{4}{3}v^{2}+
\sfrac{29}{12}v^{4})\,\mu + \sfrac{10}{9}\,\Th\,(\D_{b}v^{b})
+ 2\,v^{b}\D_{b}(\D_{c}v^{c}) + \sfrac{3}{4}\,\mu^{-1}\,
(\D_{b}\D^{b}\mu) \nonumber \\
& & \hspace{15mm} + \ 3\,(v_{b}\D_{c}\D^{[b}v^{c]})
- 3\,(\D_{[b}v_{c]}\D^{[b}v^{c]}) + 3\,(v^{b}\ndot^{c}\D_{[b}v_{c]})
\nonumber \\
& & \hspace{15mm} + \ 3\,\mu^{-1}\,(v_{b}\D^{[b}v^{c]}\D_{c}\mu)
\ ]\,v^{a} \nonumber \\
& & \hsp5 + \ v_{b}\,[\ 3\,v^{c}\D_{c}\D^{[a}v^{b]}
+ \sfrac{3}{4}\,\mu^{-1}\,\D^{b}\D^{a}\mu
+ 5\,(\D_{c}v^{c})\,\D^{[a}v^{b]} + 3\,\mu^{-1}\,(v^{c}\D_{c}\mu)\,
\D^{[a}v^{b]} \nonumber \\
& & \hspace{15mm} + \ 3\,(v_{c}\ndot^{c})\,\D^{[a}v^{b]}
+ \sfrac{3}{2}\,\D^{\lgl a}\ndot^{b\rgl}
+ \sfrac{3}{2}\,\ndot^{\lgl a}\,\ndot^{b\rgl}\ ] \nonumber \\
& & \hsp5 + \ \sfrac{1}{2}\,(\D_{b}v^{b})\,\mu^{-1}\,\D^{a}\mu
- \sfrac{3}{4}\,\D^{\lgl a}v^{b\rgl}\,\mu^{-1}\,(\D_{b}\mu)
- \sfrac{1}{4}\,\D^{[a}v^{b]}\,\mu^{-1}\,(\D_{b}\mu) \nonumber \\
& & \hsp5 + \ 3\,\D^{\lgl a}v^{b\rgl}\,\D_{[b}v_{c]}\,v^{c}
+ 3\,\D^{[a}v^{b]}\,\D_{\lgl b}v_{c\rgl}\,v^{c}
- 6\,\D^{[a}v^{b]}\,\D_{[b}v_{c]}\,v^{c} \ .
\eea
We remark that, besides $(C_{1})^{a}$ to $(C_{5})^{a}$, in
principle also $(C_{G})$ and $(C_{G})^{ab}$ given in
Eqs. (\r{covfried}) and (\r{covan3ric}) are spatial constraints
which need to be preserved along ${\bf n}$. However, the $1+3$
covariant formulation does {\em not\/} provide natural evolution
equations for the spatial curvature variables ${}^{3}\!R$ and
${}^{3}\!S_{ab}$, or, rather, the underlying connection components
relating to $h_{ab}$. To complete the constraint analysis, one
instead needs to introduce either a $1+3$ orthonormal frame (\,and
the equations deriving from the Jacobi and Ricci identities listed
in subsection \r{subsec:13onfqn} of the appendix\,), or a set of
local coordinates (\,see, e.g., Ref. \ct{vel97}\,). Experience
shows, nevertheless, that in general both $(C_{G})$ and
$(C_{G})^{ab}$ {\em are\/} preserved (\,cf., e.g.,
Refs. \ct{vanetal97} and \ct{vel97}\,).

Returning to the set (\r{c1dot}) - (\r{c5dot}), we now see that
with the choice (\r{ndot}) both $(C_{1})^{a}$ and $(C_{2})^{a}$
will be preserved along ${\bf n}$, if each of $(C_{1})^{a}$,
$(C_{2})^{a}$ and $(C_{5})^{a}$ hold on an initial spacelike
3-surfaces ${\cal T}_{0}$: $\left\{t_{0}=\mbox{const}\right\}$.
However, propagation of $(C_{4})^{a}$ generates the {\em
secondary\/} constraint $(C_{6})^{a}$.\footnote{Note that the
structure of $(C_{6})^{a}$ is quite reminiscent of the general
$\D_{b}\sig^{ab}$-constraint (Ricci identity) in that it relates
the spatial divergence of the shear $\D_{\langle a}v_{b\rangle}$ of
the peculiar velocity field $v^{a}$ to the spatial gradient of its
expansion $\D_{a}v^{a}$.}  Furthermore, it is clear that neither
$(C_{3})^{ab}$ nor $(C_{5})^{a}$ can be made to be preserved along
${\bf n}$ for a {\em single\/} value of $\alpha$, so they also
generate secondary constraints. Hence, exact quasi-Newtonian
cosmologies with dust matter source field and a lapse function
evolution given by Eq. (\r{ndot}) are not consistent at the first
derivative level as regards propagation of the constraints along
${\bf n}$ --- new conditions arise. The question is whether
consistency can be attained at a higher derivative level, or with a
prescription for $\dot{N}$ alternative to Eq. (\r{ndot}); however,
in view of the complexity which arises in the situation just
discussed this seems doubtful except in very special cases. Indeed,
initially our hope was that this project could lead to a
substantial class of exact non-linear general relativistic
generalisations of the Zel'dovich approximation method for
gravitational instability scenarios; however, considering the
complexity of the consistency equations, and noting the physical
restrictions implied by excluding all gravitational radiation
phenomena, we now believe there exist only very few consistent
quasi-Newtonian cosmologies with dust matter source field.

We have been unable to resolve the generic consistency question by
either $1+3$ covariant or orthonormal frame methods. Thus, we end
up with the open issue:
\begin{quotation}
{\bf Problem 3}: Determine (a) the set of cases where the
constraint equations for a quasi-Newtonian cosmologies with dust
matter source field are consistent with the time evolution
equations, and (b) the related set of consistent quasi-Newtonian
cosmologies.
\end{quotation}
In the following section we derive some of the consequences that
arise from imposing particular dynamical restrictions on
quasi-Newtonian cosmologies with dust matter source
field. Thereafter, in section \r{sec:qnflrwlin}, we linearise the
dynamical equations about a non-comoving FLRW background.

\section{Constraint analysis in restricted cases}
\l{sec:qnconsres}
We have examined the reduced set of $1+3$ covariant dynamical
equations when particular geometrical restrictions are placed on
the quasi-Newtonian cosmologies with dust matter source field, to
see if we can solve the problem in those cases. In particular, we
have examined quasi-Newtonian cosmologies with (a) zero spatial
rotation of $v^{a}$, (b) zero spatial divergence of $v^{a}$, (c)
LRS spacetime symmetry. In each case we find a more manageable
reduced set of equations, but nevertheless have been unable to
resolve the problem.

Where we have made more progress is in the cases of quasi-Newtonian
cosmologies with (a) zero electric Weyl curvature (the conformally
flat case), (b) the comoving case, and (c) zero relativistic
acceleration. We now consider those cases in turn.

\subsection{Zero $E_{ab}$ --- FLRW in a non-comoving frame}
Specialising to zero electric Weyl curvature with respect to ${\bf
n}$, it follows from Eq. (\r{weyln}) that
\be
\l{cflat}
0 = E_{ab} = H_{ab} \hsp5 \Leftrightarrow \hsp5 C_{abcd} = 0 \ .
\ee
Now for $\mu \neq 0$ and $\Th \neq 0$ it is well known that the
{\em only\/} conformally flat non-static dust cosmologies belong to
the FLRW family (\,see, e.g.,
Refs. \ct{ehl61,ell71,ksmh80,pee69}\,). Thus, the expanding
non-empty cosmological models of this kind are just the FLRW
geometries as seen by a non-comoving Eulerian observer in a
Newtonian-like rest frame (in this case these frames are not
unique, because of the vanishing Weyl curvature, cf. the
discussions in subsections \r{subsec:qnunique} and
\r{subsubsec:qnflrwtilt}).

{}From Eq. (\r{covdive}) we then find that $(C_{4})^{a}$ becomes
\be
\l{ncflrw1}
(C_{4})^{a} = - \sfrac{1}{3}\,(1+\sfrac{1}{2}v^{2})\,\D^{a}\mu
+ \sfrac{1}{3}\,[\ \Th + \sfrac{3}{2}\,\mu^{-1}\,(v^{b}\D_{b}\mu)
+ \sfrac{5}{3}\,(\D_{b}v^{b})\ ]\,\mu\,v^{a}
+ \sfrac{1}{6}\,\mu\,v_{b}\,[\ \D^{\lgl a}v^{b\rgl}
- 5\,\D^{[a}v^{b]}\ ] \ ,
\ee
(essentially: the equation $\D_{b}E^{ab} = 0$), which can be
treated either as an equation for $\D_{a}\mu$ (determining the
change of $\mu$ along the direction $v^{a}$ in ${\cal T}$:
$\left\{t=\mbox{const}\right\}$),\footnote{Even though this is a
FLRW model, this will be non-zero if $v^{a} \neq 0$ because then we
are in a tilted frame (\,cf. Ref. \ct{kinell73}\,).} or for
$\D_{a}v_{b}$. Then, Eq. (\r{covedot}) provides the new constraint
(essentially: the equation $\dot{E}^{\langle ab\rangle} = 0$)
\bea
(C)^{ab} & := & \sfrac{2}{3}\,(1-\sfrac{1}{4}v^{2})\,\Th\,\mu\,
v^{\lgl a}\,v^{b\rgl} + \sfrac{1}{2}\,(v^{c}\D_{c}\mu)\,
v^{\lgl a}\,v^{b\rgl} - \sfrac{1}{2}\,v^{\lgl a}\,\D^{b\rgl}\mu
\nonumber \\
& & \hsp5 + \ \sfrac{5}{6}\,(\D_{c}v^{c})\,\mu\,
v^{\lgl a}\,v^{b\rgl}
- \sfrac{1}{2}\,\mu\,\D^{\lgl a}v^{b\rgl}
+ \mu\,v_{c}\,v^{\lgl a}\,\D^{\lgl b\rgl}v^{c\rgl}
+ \mu\,v_{c}\,v^{\lgl a}\,\D^{[b\rgl}v^{c]} \ ,
\eea
which is also an equation for $\D_{a}\mu$ or $\D_{a}v_{b}$. We will
discuss the non-comoving FLRW case in linearised form in detail in
section \r{sec:qnflrwlin} below.

\subsection{Zero $v^{a}$ --- The comoving case}
\l{subsec:qncomov}
If we impose the comoving condition $v^{a} = 0$, it immediately
follows from Eq. (\r{covvdot}) that the relativistic acceleration
vanishes:\footnote{We have imposed the condition that the matter is
pressure-free (and so moves geodesically), and now that matter is
moving along the integral curves of ${\bf n}$; hence, these must be
geodesic.}
\be
\l{comoving}
v^{a} = 0 \hsp5 \Rightarrow \hsp5 \ndot^{a} = 0 \ .
\ee
Then the equations simplify dramatically because by
Eq. (\r{covsigdot})
\be
E_{ab} = 0 \ .
\ee
Thus, as condition (\r{cflat}) is satisfied, the only cosmologies
satisfying Eq. (\r{comoving}) are the FLRW models in their standard
(comoving) frame.

In detail: as $\dot{n}^{a} = 0$, we have by Eq. (\r{acc}) that $N =
N(t)$; the relativistic acceleration scalar potential is spatially
constant,\footnote{That is, it is constant in the preferred
spacelike 3-surfaces ${\cal T}$: $\left\{t=\mbox{const}\right\}$
orthogonal to ${\bf n}$.} and we can choose $N = 1$ without loss of
generality such that $t$ in Eq. (\r{qnds2}) then measures proper
time of physical dimension $\lgth$ (\,see, e.g.,
Ref. \ct{ell71}\,). From $(C_{1})^{a} = 0$, $\D_{a}\Th = 0
\Rightarrow$ we can choose local coordinates so that $S = S(t)$ in
Eq. (\r{qnds2}). From $(C_{2})^{a} = 0$, $\D_{a} \mu = 0$, so
Eq. (\r{covfried}) $\Rightarrow \D_{a}{}^{3}\!R = 0$ and
Eq. (\r{covan3ric}) $\Rightarrow {}^{3}\!S_{ab} = 0$. Hence, these
cosmologies are spatially homogeneous (\,i.e., $\mu = \mu (t)$,
$\Th =\Th(t)$, ${}^{3}\!R ={}^{3}\!R(t)$\,) with orthogonal
spacelike 3-surfaces with 3-metric $h_{ab}$ of {\em constant\/}
curvature $K(t) := {}^{3}\!R/6$. Thus, the 3-metric
$f_{\alpha\beta}(x^{\gamma})$ in Eq. (\r{qnds2}) can be chosen in
one of the standard forms for a 3-space of constant dimensionless
scalar curvature $k$ where $k$ can be normalised to one of $\{\,0,
\,\pm\,1\,\}$ by suitably renormalising $S(t)$ (\,see, e.g.,
Ref. \ct{ell87}\,). The primary constraint equations are then all
identically satisfied, and the remaining non-trivial propagation
equations are Eqs. (\r{covthdot}) and (\r{covmudot}), with a first
integral given by Eq. (\r{covfried}), provided $K(t)=k/S^{2}(t)$,
which will indeed hold.

It is fundamentally important for what follows later, that by
condition $(C_{1})^{a} = 0$ the comoving case is precisely that
case where the spatial gradient of the expansion rate is zero:
\be
\l{vel}
v^{a} = 0 \hsp5 \Leftrightarrow \hsp5
q^{a} = 0 \hsp5 \Leftrightarrow \hsp5 \D^{a}\Th = 0 \ .
\ee
Thus, we note that any of these conditions imply a comoving FLRW
subcase of our quasi-Newtonian setting.

\subsection{Zero $\dot{n}^{a}$ --- The geodesic case}
\l{subsec:qngeod}
When the relativistic acceleration of the shearfree normals ${\bf
n}$ vanishes, again
\be
\l{zero_acc}
\ndot^{a} = 0 \hsp5 \Leftrightarrow \hsp5 N = N(t) \ ;
\ee
the preferred (Newtonian-like) Eulerian observers are in free fall,
and so experience no effective gravitational field. The canonical
choice for the lapse function is $N = 1$ (cf. subsection
\r{subsec:qncomov}). Indeed, we have a zero-acceleration
quasi-Newtonian cosmologies if and only if there are local
coordinates in which the line element reduces to Eq. (\r{qnds2})
with $N = 1$. Examples are Minkowski spacetime and the family of
FLRW models. The question is whether there are other
quasi-Newtonian cosmologies satisfying Eq. (\r{zero_acc}). For the
conditions specified by the latter, Eq. (\r{accdot}), which does
{\em not\/} assume a particular form for $\dot{N}$, only leads to
an identity and so provides no new information. On the basis of the
Ansatz (\r{ndot}), however, we find from Eq. (\r{covaccdot}) that
for $\alpha \neq 0$, $\mu \neq 0$ and $(C_{1})^{a}$ satisfied,
\be
\ndot^{a} = 0 \hsp5 \Rightarrow \hsp5 v^{a} = 0 \ ;
\ee
this is just the comoving case just discussed in subsection
\r{subsec:qncomov}.

\section{Linearised perturbations of a FLRW dust cosmology}
\l{sec:qnflrwlin}
When discussing the evolution of {\em linear\/} spatially
inhomogeneous deviations from a FLRW spacetime geometry, it is a
common feature of many papers to introduce local coordinates and an
explicit Ansatz for the infinitesimal line element (\,see, e.g.,
Refs. \ct{bar80,ber92,pee80}\,). However, in doing so important
physical aspects of the dynamics of the setting often become rather
obscured. We claim that it makes for clarity to work in $1+3$
covariant terms for as long as possible. In this section we discuss
the important issue of a consistent Newtonian limit of
FLRW-linearised cosmologies, once we have provided a $1+3$
covariant formulation of the setting at hand. For a comparison
discussion, we first briefly review a recent treatment given by
Bertschinger of the evolution of linear energy density
inhomogeneities (i.e., so-called `scalar modes') as induced by
matter in {\em non\/}-relativistic motion \ct{ber92}. The reason
for referring to his work is that the timelike reference congruence
he chooses is of the Newtonian-like kind, i.e., it is irrotational
and shearfree. Other well-known discussions such as, e.g., that
given by Peebles in Ref. \ct{pee80} pick the timelike reference
congruence to be irrotational and geodesic instead (\,in Bardeen's
terminology, the latter choice provides `synchronous' gauge
conditions \ct{bar80}\,). We briefly refer to them at the end.

\subsection{Bertschinger's method}
Bertschinger chooses what he calls `conformal Newtonian' gauge
conditions \ct{ber92}. In Bardeen's terminology these are
`longitudinal' gauge conditions, where the timelike normals ${\bf
n}$ to the spacelike 3-surfaces ${\cal T}$:
$\left\{t=\mbox{const}\right\}$ are shearfree \ct{bar80}, and it is
assumed that the matter perturbations do {\em not\/} produce any
anisotropic pressures. The discussion in Ref. \ct{ber92} then
focuses on scalar linear spatially inhomogeneous
deviations. Choosing a set of Cartesian spatial coordinates
$\vec{x}:=(x,y,z)$ comoving with ${\bf n}$ on a spatially flat FLRW
background as well as a conformally rescaled time variable $\eta$
defined by\footnote{If, contrary to the time coordinate in
Eq. (\r{qnds2}), $t$ here denotes dimensional proper time, then
$\eta$ is dimensionless.}
\be
\l{conft}
\frac{d}{dt} := \frac{1}{S}\,\frac{d}{d\eta}
\hsp5 \Rightarrow \hsp5
\frac{d^{2}}{dt^{2}} = \frac{1}{S^{2}}
\,\left[\ \frac{d^{2}}{d\eta^{2}}
- \frac{1}{S}\,\frac{dS}{d\eta}\,\frac{d}{d\eta}\ \right] \ ,
\ee
it is stated that only one (scalar) variable, $\Phi$, is needed to
express the deviations in the background line element. Though no
explicit explanation is given, this follows from the assumption of
{\em zero\/} anisotropic pressure perturbations in combination with
part of the linearised EFE \ct{muketal92}; we give the relevant
details in subsection \r{subsec:flrw0p} of the appendix. One so
obtains the line element reduced to the simple form
\be
\l{berds2}
ds^{2} = S^{2}(\eta)\,[\ -\,(1+2\Phi)\,d\eta^{2}
+ (1-2\Phi)\,(dx^{2}+dy^{2}+dz^{2})\ ] \ .
\ee
{\em No\/} further statement is made in Ref. \ct{ber92} as to the
$\eta$-dependence of $\Phi $.

In the local coordinates employed, the proper peculiar velocity of
a matter particle is given by $\vec{v} = d\vec{x}/d\eta$. It should
here be stressed that in the Eulerian description adopted by
Bertschinger the peculiar velocity ``is defined as the average
momentum per unit mass of the particles in the vicinity of a given
Eulerian position'' \ct{ber92}, i.e., in terms of the {\em momentum
density\/} of the non-relativistically moving matter.
Gravitational effects generated by the momentum density, however,
are said to be neglected. Dynamical equations are now derived from
the remaining linearised EFE on the basis of the assumptions that
(1) the observable part of the Universe is nearly FLRW on scales of
the present day Hubble radius, (2) $|\,\Phi\,| \ll 1$ and $v^{2}
\ll 1$, (3) only spatial inhomogeneities on scales much less than
the present day Hubble radius are considered, and (4) the
deviations in the energy density are dominated by
non-relativistically moving matter only, so no anisotropic
pressures need to be accounted for \ct{ber92}. For dust, the set
presented in Ref. \ct{ber92} to evolve and constrain the deviations
corresponds to
\bea
\l{ber1}
\d\mu' & = & -\,3\,\frac{S'}{S}\,\d\mu
- \mu_{m}\,\nabla\cdot\vec{v} - \nabla\cdot(\d\mu\,\vec{v}) \\
\l{ber2}
\vec{v}\,' & = & -\,\frac{S'}{S}\,\vec{v}
- (\vec{v}\cdot\nabla)\,\vec{v} - \nabla\Phi \\
\l{ber3}
0 & = & \nabla\cdot\nabla\Phi - \sfrac{1}{2}\,S^{2}\,\d\mu \ ;
\eea
the prime denotes a partial derivative with respect to conformal
time $\eta$, the $\nabla$-operator acts in terms of the comoving
Cartesian coordinates $\vec{x}$, and the definition $\d\mu :=\mu -
\mu_{m}$ was used ($\mu_{m}$ being the mean total energy density of
the non-relativistic matter). Equation (\r{ber1}) is the continuity
equation representing conservation of energy (density),
Eq. (\r{ber2}) is the Euler equation representing conservation of
momentum (density), while Eq. (\r{ber3}) is the PFE of Newtonian
gravitation. In view of the Ansatz (\r{berds2}) for a spatially
flat background, the PFE describes how matter inhomogeneities
induce the irregularities in the almost-FLRW line element.

Keeping only terms of first-order smallness, Bertschinger derives
from the set (\r{ber1}) - (\r{ber3}) the standard second-order
linear ODE \ct{ber92,pee80}
\be
\l{ber2ndoode}
0 = \d'' + \frac{S'}{S}\,\d' - \sfrac{1}{2}\,S^{2}\,\mu_{m}\,\d
\ee
for the {\em dimensionless density contrast\/} $\d :=
\d\mu/\mu_{m}$. This is, however, a highly gauge and coordinate
dependent procedure. By contrast, a corresponding equation can be
derived in a more gauge-independent and covariant way by using
geometrical variables and their dynamical equations. In the
comoving case, this reproduces exactly the equation
(\r{ber2ndoode}) (\,see Ref. \ct{ellbru89}\,); however, in the
non-comoving case corresponding to the situation discussed here,
generically it gives a different result, as we now explain.

\subsection{Covariant and gauge-invariant formulation}
We contrast the description given by Bertschinger with a {\em
non\/}-comoving $1+3$ covariant and gauge-invariant formulation of
linearised perturbations of FLRW models in the spirit of Ellis and
Bruni \ct{ellbru89}, but in the quasi-Newtonian cosmology
context. That is, (\,consistent with Eq. (\r{berds2})\,) we assume
the time slicing of $\left(\,{\cal M},\,{\bf g},\,{\bf n}\,\right)$
provided by the spacelike 3-surfaces ${\cal T}$:
$\left\{t=\mbox{const}\right\}$ with shearfree normals ${\bf n}$ to
have an approximately FLRW geometry, such that the quantities
$v^{a}$ and $\ndot^{a}$ can be regarded as (gauge-invariant)
deviations of {\em first\/}-order smallness. As we consider
non-relativistically moving matter only, we {\em do\/} account for
its momentum density $q^{a}$, which is {\em linear\/} in $v^{a}$,
but {\em neglect\/} its isotropic and anisotropic pressures, $p$
and $\pi_{ab}$, as they are of second-order smallness in $v^{a}$
(\,cf. Eqs. (\r{mat})\,). The difference to the Ellis--Bruni
treatment is that here we are {\em not\/} using a comoving choice
of the timelike reference congruence, thus, the procedure is
frame-independent iff the Newtonian-like timelike reference
congruence ${\bf n}$ is unique. We expect this to be the case in
general, but, as discussed in subsection \r{subsec:qnunique} above,
there may be some freedom in its choice in specific restricted
geometrical cases. In particular, we can view a strictly FLRW model
from a tilted (non-comoving) reference frame. We will come back to
the aspect of factoring out the pure non-comoving FLRW subcase,
once we have investigated the general properties of the
FLRW-linearised quasi-Newtonian dynamical equations.

\subsubsection{The lapse function in almost-FLRW models}
Given this context, we now return to the question as to how to
evolve the lapse function $N$ along the integral curves of our
timelike reference congruence ${\bf n}$. The standard literature
often deals with this aspect by merely stating (though generally
not providing explicit mathematical relations; however, see, e.g.,
Refs. \ct{ehlsch93,holwal97,seietal94} for cases where the
background is expanding, and Ref. \ct{wal84} where it is not) that
the non-relativistic peculiar motion of the matter induces only
`slowly' evolving variations in the metric structure of
$\left(\,{\cal M},\,{\bf g},\,{\bf n}\,\right)$. One way to
interpret statements of this kind is to take $\dot{N} = 0$, but
other options may also be taken into account. A number of possible
ways of evolving the lapse function $N$ along ${\bf n}$ are
contained in the Ansatz (\r{ndot}), which we continue to use in the
subsequent analysis, leading to the evolution equation
(\r{covaccdot}) for $\ndot^{a}$. Besides the `harmonic gauge' given
by $\alpha = 1$, this Ansatz also allows for the special
subcases:\enl

(i) $\alpha = -\,\sfrac{1}{3} \Rightarrow (SN)\,{}\dot{} = 0$; the
dimensionless comoving (with the Eulerian observers) lapse function
$(SN)$ is covariantly constant along ${\bf n}$; this choice is
suggested from computing to first order the $1+3$ ONF commutation
functions associated with the line element (\r{berds2}): as can be
derived from the relations given in subsection \r{subsec:flrw0p} of
the appendix one obtains $\ndot^{\alpha} = a^{\alpha}$, $0 =
\sig_{\alpha\beta} = \om^{\alpha} = n_{\alpha\beta} =
\Omega^{\alpha}$, and so the evolution equation for
$\ndot^{\alpha}$ is given by Eq. (\r{jac1}) in subsection
\r{subsec:13onfqn}.\enl

(ii) $\alpha = 0 \Rightarrow \dot{N} = 0$; as mentioned above, here
the lapse function $N$ itself is covariantly constant along ${\bf
n}$;\enl

(iii) $\alpha = \sfrac{1}{3}$; provided one makes the
identification $N = S\,(1+2\Phi)^{1/2}$,\footnote{At this point,
one faces a problem with the physical dimension of the time
coordinates involved. While our $t$ in Eq. (\r{qnds2}) is
dimensionless, letting (as is standard in the ADM $3+1$ formalism)
$N$ carry the dimension of the line element, in other works such as
Refs. \ct{muketal92} and \ct{holwal97} $t$ is dimensional proper
time (and $N = 1$). This prevents an unambiguous identification of
$N$ with variables such as $S$ and $\Phi$, once a (dimensionless)
conformal time coordinate $\eta$ is introduced.}  this value is
implied by the analysis of {\em growing\/} mode `adiabatic'
perturbations in spatially flat FLRW backgrounds given by Mukhanov
et al \ct{muketal92}, p225, where (in their notation) $\Phi' =
0$.\enl

\subsubsection{Evolution of constraints}
Besides $v^{a}$ and $\ndot^{a}$, also the $1+3$ covariant
quantities $X_{a} := \D_{a}\mu$ and $Z_{a} := \D_{a}\Th$ encode
deviations of first-order smallness from an exact FLRW geometry of
our time slicing; indeed we will regard $X_{a}$ and $Z_{a}$ as our
prime perturbation variables. The evolution equations for these
variables are\enl

\noindent
{\em FLRW-linearised perturbation evolution equations\/}:
\bea
\l{xdot}
\dot{X}^{\lgl a\rgl} & = & -\,\sfrac{4}{3}\,\Th\,X^{a}
- \mu\,Z^{a} - \mu\,\D^{a}(\D_{b}v^{b}) - \Th\,\mu\,\ndot^{a} \\
\l{zdot}
\dot{Z}^{\lgl a\rgl} & = & -\,\Th\,Z^{a} - \sfrac{1}{2}\,X^{a}
+ \D^{a}(\D_{b}\ndot^{b}) - \sfrac{1}{3}\,\Th^{2}\,\ndot^{a}
- \sfrac{1}{2}\,\mu\,\ndot^{a}
\ .
\eea
Note that these equations differ from Eqs. (49) and (50) for the
comoving Lagrangean ${\bf \tilde{u}}$-frame in Ellis and Bruni
\ct{ellbru89} by the presence of the second-order derivative terms
as well as the terms algebraic in $\ndot^{a}$. Of importance in the
following will also be the first-order evolution equations for the
spatial divergences of both $v^{a}$ and $\ndot^{a}$, given by
\bea
(\D_{a}v^{a})\,\dot{} & = & -\,\sfrac{2}{3}\,\Th\,\D_{a}v^{a}
- \D_{a}\ndot^{a} \\
(\D_{a}\ndot^{a})\,\dot{} & = & (\alpha-\sfrac{2}{3})\,\Th\,
\D_{a}\ndot^{a} + \alpha\,\sfrac{3}{2}\,\mu\,\D_{a}v^{a}
+ \alpha\,\D_{a}(C_{1})^{a} \ ,
\eea
respectively.

In terms of $X_{a}$ and $Z_{a}$ the constraint equations
(\r{covdivsig}) - (\r{covdive}), (\r{covc5}) and (\r{covc6}) now
assume the first-order forms \enl

\noindent
{\em FLRW-linearised constraint equations\/}:
\bea
\l{linc1}
0 & = & (C_{1})^{a} \ = \ \ Z^{a} - \sfrac{3}{2}\,\mu\,v^{a} \\
\l{linc2}
0 & = & (C_{2})^{a} \ = \ \eps^{abc}\,\D_{b}\ndot_{c} \\
\l{linc3}
0 & = & (C_{3})^{ab} \ = \ E^{ab} - \D^{\lgl a}\ndot^{b\rgl} \\
\l{linc4}
0 & = & (C_{4})^{a} \ = \ \D_{b}E^{ab} - \sfrac{1}{3}\,X^{a}
+ \sfrac{1}{3}\,\Th\,\mu\,v^{a} \\
\l{linc5}
0 & = & (C_{5})^{a} \ = \ \ \D^{a}(\D_{b}\ndot^{b})
- \sfrac{1}{2}\,X^{a} + \sfrac{1}{2}\,\Th\,\mu\,v^{a}
- \sfrac{1}{3}\,\Th^{2}\,\ndot^{a} + \mu\,\ndot^{a} \\
\l{linc6}
0 & = & (C_{6})^{a} \ = \ \D^{a}(\D_{b}v^{b})
- \sfrac{3}{2}\,\D_{b}\D^{\lgl a}v^{b\rgl}
- \sfrac{1}{3}\,\Th^{2}\,v^{a} + \mu\,v^{a} \ .
\eea
The momentum density of the moving matter $q^{a} = \mu\,v^{a}$
induces a spatial gradient $Z^{a}$ in the expansion rate of the
normals ${\bf n}$ by $(C_{1})^{a}$. Also note that in the given
situation $E_{ab} = \D_{\lgl a}\ndot_{b\rgl}$, provided
$(C_{3})^{ab}$ is satisfied and preserved along ${\bf n}$; to first
order the electric Weyl curvature of $\left(\,{\cal M}, \,{\bf g},
\,{\bf n}\,\right)$ is completely determined by the distortion of
$\ndot^{a}$. This relation reflects the connection between the
tidal field and the gravitational scalar potential in the proper
Newtonian theory \ct{ell71}. We remark that in first-order
approximation the 3-Cotton--York tensor of Eq. (\r{covan3cy}) is
zero.

We now find that the set of constraints (\r{linc1}) - (\r{linc6})
evolves to first order according to
\bea
\l{linc1dot}
(\dot{C}_{1})^{\lgl a\rgl} & = & -\,\Th\,(C_{1})^{a}
+ (C_{5})^{a} \\
\l{linc2dot}
(\dot{C}_{2})^{\lgl a\rgl} & = & (\alpha-\sfrac{2}{3})\,
\Th\,(C_{2})^{a} \\
\l{linc3dot}
(\dot{C}_{3})^{\lgl ab\rgl} & = & -\,\Th\,(C_{3})^{ab}
- \alpha\,\D^{\lgl a}(C_{1})^{b\rgl}
- (\alpha+\sfrac{1}{3})\,[\ \Th\,\D^{\lgl a}\ndot^{b\rgl}
+ \sfrac{3}{2}\,\mu\,\D^{\lgl a}v^{b\rgl}\ ] \\
\l{linc4dot}
(\dot{C}_{4})^{\lgl a\rgl} & = & -\,\sfrac{4}{3}\,\Th\,(C_{4})^{a}
+ \sfrac{1}{3}\,\mu\,(C_{1})^{a} + \sfrac{1}{3}\,\mu\,(C_{6})^{a}
\\
\l{linc5dot}
(\dot{C}_{5})^{\lgl a\rgl} & = & (\alpha-1)\,\Th\,(C_{5})^{a}
+ \alpha\,\D^{a}(C_{A})
- \alpha\,\sfrac{1}{3}\,\Th^{2}\,(C_{1})^{a}
+ (\alpha+\sfrac{1}{2})\,\mu\,(C_{1})^{a} \nonumber \\
& & \hsp5 + \ \sfrac{3}{2}\,(\alpha+\sfrac{1}{3})\,\mu\,
[\ \D^{a}(\D_{b}v^{b}) + \sfrac{1}{3}\,\Th\,\mu^{-1}\,X^{a}
- \sfrac{2}{3}\,\Th^{2}\,v^{a} + \mu\,v^{a}\ ] \\
\l{linc6dot}
(\dot{C}_{6})^{\lgl a\rgl} & = & -\,\Th\,(C_{6})^{a}
- \sfrac{3}{2}\,\D_{b}(C_{3})^{ab} + \sfrac{3}{2}\,(C_{4})^{a}
- (C_{5})^{a} \\
\l{lincadot}
(\dot{C}_{A}) & = & -\,\sfrac{4}{3}\,\Th\,(C_{A})
+ \D_{a}(C_{5})^{a} \ ,
\eea
where we have introduced the auxiliary constraint $(C_{A}) :=
\D_{a}(C_{1})^{a}$ in order to bring the evolution system
(\r{linc1dot}) - (\r{lincadot}) into a form suitable for
application of the Cauchy--Kowalewskaya theorem on existence and
uniqueness of solutions for sets of {\em analytic\/} initial data
(\,see, e.g., Ref. \ct{wal84}\,). As can be immediately seen from
Eqs. (\r{linc3dot}) and (\r{linc5dot}), we obtain the remarkable
result that {\em the set of constraints (\r{linc1}) - (\r{linc6})
evolves consistently (without restrictions on the background model)
for the special parameter value of $\alpha = -\,\sfrac{1}{3}$ in
Eq. (\r{ndot})\/}. As mentioned before, this is the situation when
the dimensionless comoving lapse function $(SN)$ is covariantly
constant along ${\bf n}$.

If, however, $\alpha \neq -\,\sfrac{1}{3}$, {\em further\/}
constraints arise, namely
\bea
\l{linc7}
0 & = & (C_{7})^{ab} \ := \ \Th\,\D^{\lgl a}\ndot^{b\rgl}
+ \sfrac{3}{2}\,\mu\,\D^{\lgl a}v^{b\rgl} \\
\l{linc8}
0 & = & (C_{8})^{ab} \ := \ \D^{\lgl a}\ndot^{b\rgl}
+ \sfrac{1}{2}\,\Th\,\D^{\lgl a}v^{b\rgl} \\
\l{linc9}
0 & = & (C_{9})^{ab} \ := \ [\ (C_{G}) - {}^{3}\!R\ ]
\,\D^{\lgl a}v^{b\rgl} \ ,
\eea
the first arising from Eq. (\r{linc3dot}), and the latter two are
each the condition resulting from consistency of the previous
one. Indeed, these propagate along ${\bf n}$ according to
\bea
\l{linc7dot}
(\dot{C}_{7})^{\lgl ab\rgl} & = & (\alpha-1)\,\Th\,(C_{7})^{ab}
+ \alpha\,\Th\,\D^{\lgl a}(C_{1})^{b\rgl} - 2\,\mu\,(C_{8})^{ab}
\\
\l{linc8dot}
(\dot{C}_{8})^{\lgl ab\rgl} & = & (\alpha-\sfrac{7}{6})\,\Th\,
(C_{8})^{ab} + \alpha\,\D^{\lgl a}(C_{1})^{b\rgl}
- \sfrac{3}{2}\,(\alpha-\sfrac{1}{6})\,(C_{9})^{ab} \\
\l{linc9dot}
(\dot{C}_{9})^{\lgl ab\rgl} & = & -\,\sfrac{5}{6}\,\Th\,
(C_{9})^{ab} - \sfrac{1}{2}\,[\ (C_{G}) - {}^{3}\!R\ ]\,
(C_{8})^{ab} \ .
\eea
Finally, one can show that for $\alpha \neq -\,\sfrac{1}{3}$ the
non-vanishing source term in the square bracket on the right-hand
side of Eq. (\r{linc5dot}) can be expressed in terms of the other
constraints, provided that ${}^{3}\!R = 0$:
\bea
(C_{10})^{a} & := & \D_{a}(\D_{b}v^{b}) + \sfrac{1}{3}\,\Th\,
\mu^{-1}\,\D^{a}\mu - \sfrac{2}{3}\,\Th^{2}\,v^{a} + \mu\,v^{a} \\
& = & (C_{6})^{a} + \Th\,\mu^{-1}\,[\ \D_{b}(C_{3})^{ab}
- (C_{4})^{a} + \D_{b}(C_{8})^{ab}\ ]
- \sfrac{3}{4}\,[\ (C_{G}) - {}^{3}\!R\ ]\,
\D_{b}\D^{\lgl a}v^{b\rgl} \ .
\eea
Hence, for $\mu \neq 0$ and $\Th \neq 0$ (and assuming that
$(C_{G})$ is satisfied and preserved along ${\bf n}$),
Eqs. (\r{linc7}) - (\r{linc9}) and (\r{linc3}) demand for $\alpha
\neq -\,\sfrac{1}{3}$ that
\be
0 = {}^{3}\!R\ \D_{\lgl a}v_{b\rgl} \hsp5
\Rightarrow \hsp5 0 = {}^{3}\!R\,E_{ab} \ ;
\ee
with this choice of evolution for $N$ a complete and
self-consistent set of variables and equations describing
linearised scalar perturbations of FLRW dust cosmologies from a
non-comoving point of view requires either
\begin{itemize}

\item a spatially flat background time slicing, ${}^3\!R=0$, or

\item a peculiar velocity field $v^{a}$ with (to first order)
vanishing shear, $\D_{\langle a}v_{b\rangle } = 0$. However, this
is just the conformally flat case which we will further discuss
below; the inhomogeneities are pure gauge in this case.

\end{itemize}
Thus, the conclusion is that {\em there are two consistent
inhomogeneous FLRW-linearised cases\/}: ${}^{3}\!R \neq 0\ ${\em
and\/} $\alpha = -\,\sfrac{1}{3}$, {\em or\/} ${}^{3}\!R = 0$ {\em
and no restriction on} $\alpha$. {\em In each case the set of
resulting equations is then consistent, but in the latter case
there are the two extra constraints (\r{linc7}) and (\r{linc8})
that must be satisfied, as well as Eqs. (\r{linc1}) - (\r{linc6}),
which must be satisfied in all cases\/}.

\subsubsection{Evolution of spatially inhomogeneous deviations}
It is now convenient to define scaled covariant
and gauge-invariant perturbation variables by
\ct{ellbru89,bruetal92}
\be
\d^{a} := \frac{S X^{a}}{\mu} \ , \hsp5 {\cal Z}^{a} := S Z^{a}
\ ,
\ee
the first being the dimensionless comoving (with the Eulerian
observers) fractional total energy density gradient, the second the
comoving rate of expansion gradient. The purely {\em scalar\/}
content of these variables is obtained by taking their magnitudes,
or, essentially equivalently, by constructing their comoving
spatial divergences \ct{elletal90,bruetal92}:
\be
\d := S\D_{a}\d^{a} \ , \hsp5 {\cal Z} := S\D_{a}{\cal Z}^{a} \ .
\ee
In terms of the latter, Eqs. (\r{xdot}) and (\r{zdot}) convert
into
\bea
\l{deldot}
\dot{\d} & = & -\,{\cal Z} - S\D_{a}[\,S\D^{a}(\D_{b}v^{b})\,]
- \Th\,S\D_{a}(S\ndot^{a}) \\
\l{calzdot}
\dot{{\cal Z}} & = & -\,\sfrac{2}{3}\,\Th\,{\cal Z}
- \sfrac{1}{2}\,\mu\,\d + S\D_{a}[\,S\D^{a}(\D_{b}\ndot^{b})\,]
- \sfrac{1}{3}\,\Th^{2}\,S\D_{a}(S\ndot^{a})
- \sfrac{1}{2}\,\mu\,S\D_{a}(S\ndot^{a}) \ ,
\eea
which, finally, after some more work, can be combined into a single
second-order linear ODE for $\d$ given by
\bea
\l{delddot}
\ddot{\d} + \sfrac{2}{3}\,\Th\,\dot{\d} - \sfrac{1}{2}\,\mu\,\d 
& = & - \,\alpha\,\sfrac{3}{2}\,\Th\,\mu\,S\D_{a}(Sv^{a})
- \sfrac{1}{2}\,[\ 2\,(\alpha-\sfrac{1}{3})\,\Th^{2} + (C_{G})
- {}^{3}\!R\ ]\ S\D_{a}(S\ndot^{a}) \nonumber \\
& & \hspace{15mm} - \ \alpha\,\Th\,S\D_{a}[\,S(C_{1})^{a}\,] \ .
\eea
In obtaining this relation, use was made of Eq. (\r{covfried}) and
the fact that, with Eq. (\r{covaccdot}), we have to first order
\be
\l{cmdivaccdot}
[\ S\D_{a}(S\ndot^{a})\ ]\,\dot{} = \alpha\,[\ \Th\,
S\D_{a}(S\ndot^{a}) + \sfrac{3}{2}\,\mu\,S\D_{a}(Sv^{a})
+ S\D_{a}[\,S(C_{1})^{a}\,]\ ] \ .
\ee
Equation (\r{delddot}) needs to be compared to results given in the
standard literature (\,see, e.g., Refs. \ct{bar80} and
\ct{pee80}\,), and the Bertschinger relation (\r{ber2ndoode}) given
above. With $(C_{1})^{a} $ and $(C_{G})$ satisfied initially and
preserved along ${\bf n}$, {\em extra velocity-induced terms
occur\/}. The most complex extra source term on the right-hand side
will arise when $\alpha = -\,\sfrac{1}{3}$ which, as derived above,
is the only case that allows for the possibility ${}^{3}\!R \neq
0$. However, even when ${}^{3}\!R = 0$, unless $\alpha = 0$ we have
to account in particular for the first source term on the
right-hand side of Eq. (\r{delddot}), which directly derives from
the momentum density of the moving matter.

As Roy Maartens \ct{maa98} kindly points out to us, in the special
case when $\alpha = 0$ (corresponding to $\dot{N} = 0$)
Eq. (\r{delddot}) can be brought into standard form (\,e.g.,
Eq. (11.1) in Ref. \ct{pee80}\,) by using $0 = \mu -
\sfrac{1}{3}\,\Th^{2}$ and defining a new scalar perturbation
variable $Y:=\d + 2\,S\D_{a}(S\dot{n}^{a})$; the second term then
is covariantly constant along ${\bf n}$ by
Eq. (\r{cmdivaccdot}). In this sense, computational simplicity
singles out the value $\alpha = 0$ amongst all cases with a
spatially flat background. The question is as to the physical
content of this feature, since, unless ${}^{3}\!R \neq 0$, the
value of $\alpha$ can be chosen completely arbitrarily. Solutions
to the modified evolution equation (\r{delddot}) for $\d$ can
probably only be obtained by numerical means.

\subsubsection{Tilted FLRW models}
\l{subsubsec:qnflrwtilt}
We now return to the special case considered above, where
$\D_{\langle a}v_{b\rangle } = 0$. This is just a FLRW model seen
from a non-comoving frame. Indeed, from the non-comoving
perspective, an Eulerian observer perceives spatial inhomogeneity
even if the dust matter flow is {\em exactly\/} FLRW, i.e., if
$$
0 =
\dot{\tl{u}}{}^{a} = \tl{\sig}_{ab} = \tl{\om}^{a}
\hsp5 \Rightarrow \hsp5 
0 = \tl{E}_{ab} = \tl{H}_{ab} \ ,
$$
because this is then a tilted spatially homogeneous cosmological
model \ct{kinell73}. To first order in the deviations we find that
if this tilted situation is given, then with $\mu \neq 0$ and $\Th
\neq 0$ it is characterised by:

(i) from Eq. (\r{covedot})
\be
\l{rest}
0 = \D_{\langle a}v_{b\rangle } \ ,
\ee
(this is the condition that the spacetime is conformally flat to
linear order); and

(ii) from Eq. (\r{linc4}), combined with Eqs. (\r{linc1}) and
(\r{covfried}),
\be
0 = (C_{4})^{a} = -\,\sfrac{1}{6}\,\D^{a}[\ {}^{3}\!R - (C_{G})\ ]
- \sfrac{2}{9}\,\Th\,(C_{1})^{a} \ ;
\ee
with $(C_{1})^{a}$ and $(C_{G})$ satisfied, the intrinsic 3-Ricci
curvature of the time slicing is found to be spatially
homogeneous. From Eq. (\r{linc6}) we then have
$0 = \D^{a}(\D_{b}v^{b}) + \sfrac{1}{2}\,{}^{3}\!R\,v^{a}$.

\subsection{Derivation of the limiting Newtonian-like gravitational
field equation}
\l{subsec:qnflrwlin3}
How does the difference between Eq. (\r{delddot}) and the Newtonian
form (\r{ber2ndoode}) given by Bertschinger arise? The key role in
standard treatments of FLRW-linearised matter and geometry
perturbations is played by the PFE, the central equation in the
Newtonian theory of gravitation, which, in our context, ties the
spatial inhomogeneities in the total energy density distribution to
the Laplacian of the relativistic acceleration scalar potential
$N$. The problem seems to lie in the assumed exactly Newtonian form
of this equation used by Bertschinger. We suggest that from the
point of view of internal consistency this may not be the
appropriate Newtonian-like gravitational field equation to use in
the cosmological context we consider here; that rather this
equation should be the PFE modified by inclusion of a momentum
density induced term.

Thus, the issue now is to discuss obtaining a consistent set of
quasi-Newtonian dynamical equations in a cosmological context from
the exact relativistic equations. Where can this come from?

\subsubsection{The Raychaudhuri equation}
The Raychaudhuri equation (\r{covthdot}) becomes the relativistic
generalisation of the PFE in the static case, when there is no
expansion of the timelike reference congruence. Setting $\Th = 0$
in Eq. (\r{covthdot}) with $v^{2} \ll 1$ gives the well-known
gravi-static equation of gravitational attraction
$$
\D_{a}\ndot^{a} + (\ndot_{a}\ndot^{a}) = N^{-1}\,\D_{a}\D^{a}N
= \sfrac{1}{2}\,\mu \ ,
$$
--- a direct generalisation of the Newtonian gravitational field
equation.\footnote{But generically with active gravitational mass
$(\mu+3p)$ (\,see, e.g., Whittaker \ct{whi35} or Ehlers
\ct{ehl61}\,).} The problem is that this does not work in the
generic cosmological context: in a {\em non\/}-static cosmological
spacetime this equation determines the rate of change of expansion
of the model, with the spatial divergence of the relativistic
acceleration as one of the source terms, rather than directly
relating that divergence to the matter energy density.

The Raychaudhuri equation would become the relativistic
generalisation of the PFE if we could suitably subtract out the
rate of expansion, leaving a relation between energy density
variation and the relativistic acceleration. The problem is that
the background expansion is spatially homogeneous but the real
expansion $\Th$ is not, so subtracting off the background expansion
terms is a gauge-dependent procedure. This will only give a
gauge-invariant result if $\Th$ is spatially homogeneous, i.e., if
$Z^{a} = 0$. To make this work, we would have to subtract off the
expansions terms not of a background model but of the real model;
but they are spatially varying, and so would add extra terms to the
PFE-like equation obtained. The resultant extra terms may be
negligible in specific contexts (e.g. gravitational lensing
models), but then that circumstance needs to be shown.

\subsubsection{The Friedmann equation}
The Friedmann equation for our quasi-Newtonian models takes the
form
$$
2\,\mu = {}^{3}\!R + \sfrac{2}{3}\,\Th^{2} \ .
$$
If the background FLRW model is spatially flat, then with the line
element in the form (\r{berds2}) this relation becomes
$$
2\,\mu = \frac{4}{S^{2}}\,\left[\ \Phi_{|xx}+\Phi_{|yy}+\Phi_{|zz}
\ \right] + \sfrac{2}{3}\,\Th^{2} \ .
$$
Hence, again if we could subtract off the expansion term
$\sfrac{2}{3}\,\Th^{2}$ we would have an equation of the form we
want. However, again that will not work in the generic physically
relevant case when $\Th$ is spatially inhomogenous; we can subtract
off the background expansion consistently in a position-independent
way, but not the real expansion, for that is
position-dependent. Again, this spatial dependence may be
negligible in particular circumstances, but that needs to be shown
(note that we are carrying out a consistent linearised study of the
equations; it is not true that linearisation by itself implies we
can neglect these terms).

What we really need to do is to take the spatial gradient of this
equation to get a covariant and gauge-invariant equation of the
form we want. However, we already have equations of that kind at
hand: namely, the $\D_{b}E^{ab}$-constraint (Bianchi identity) and
the compatibility condition $(C_{5})^{a}$ binding the former to the
shearfree slicing condition $(C_{3})^{ab}$; we consider this next.

\subsubsection{The $\D_{b}E^{ab}$-constraint (Bianchi identity)}
In the case of a spatially flat background, the compatibility
condition $(C_{5})^{a}$ between the $\D_{b}E^{ab}$-constraint
(Bianchi identity) $(C_{4})^{a}$ and the zero-shear condition
$(C_{3})^{ab}$ assumes from Eq. (\r{linc5}) the form
$$
\D^{a}(\D_{b}\ndot^{b}) = \sfrac{1}{2}\,X^{a}
- \sfrac{1}{2}\,\Th\,\mu\,v^{a} \ .
$$
In terms of the relativistic acceleration scalar potential $N$,
this relation can be re-written as an effective Newtonian
(PFE-like) gravitational field equation given by
$$
\D^{a}[\ \D_{b}\D^{b}N - \sfrac{1}{2}\,N\,\mu\ ]
= -\,\sfrac{1}{2}\,N\,\Th\,\mu\,v^{a} \ ; 
$$
this is vectorial and third-order because it is covariant and
gauge-invariant,\footnote{Again, its purely scalar content is
obtained from the spatial divergence of this relation.} and it has
an extra velocity part essentially resulting from the momentum
constraint equation $(C_{1})^{a} = 0$. We believe that any other
approach that consistently derives a {\em scalar second-order
PFE-like equation\/} in the current context will also lead to such
an extra term.

\subsubsection{Neglecting the momentum density}
If (\,as, e.g., in Ref. \ct{holwal97}\,) one uses a version of any
of these equations without the term proportional to $q^{a} =
\mu\,v^{a}$, then one has neglected the first-order momentum
density due to the non-relativistic peculiar motion of the matter.
While this term is small, the implication, via $(C_{1})^{a} = 0$,
is that one has also neglected the spatial gradient $Z^{a}$ of the
expansion rate $\Th$ of ${\bf n}$; for example, the set (\r{ber1})
- (\r{ber3}) {\em neglects\/} this $(0\alpha )$-EFE,\footnote{The
$(0\alpha )$-EFE is equivalent to the $\D_{b}\sig^{ab}$-constraint
(Ricci identity) of the $1+3$ covariant dynamical equations.} which
links the matter momentum density --- {\em linear\/} in $\vec{v}$
--- to mixed time--space derivatives of $\Phi$
(\,cf. Ref. \ct{holwal97} and subsection \r{subsec:flrw0p} of the
appendix\,).

To investigate this, we now inquire into the viability of imposing
the {\em extra\/} assumption
$$
q^{a} = \mu\,v^{a} = 0 \hsp5 \Rightarrow \hsp5 Z^{a} = 0 \ ;
$$
thus, we set $q^{a}$ to zero throughout, {\em implying\/}, as $\mu
\neq 0$, that $v^{a}$ vanishes. We here find ourselves in a
slightly delicate situation, as, e.g, Bertschinger \ct{ber92}
neglects $q^{a}$ but not $v^{a}$ itself (part of his dynamical set
is an evolution equation for $\vec{v}$). Note that, to first order,
we have, e.g., $\D_{\lgl a}q_{b\rgl} = \mu\,\D_{\lgl a}v_{b\rgl}$
and $\D_{a}q^{a} = \mu\,\D_{a}v^{a}$.

{}From the gauge problem point of view, our extra assumption
amounts to choosing a family of spacelike 3-surfaces ${\cal T}$:
$\left\{t=\mbox{const}\right\}$ with timelike normals ${\bf n}$
which, {\em simultaneously\/}, are shearfree and have spatially
constant rate of expansion \ct{bar80}.  However, for $\mu \neq 0$
and $\Th \neq 0$, it now follows that if $\D_{\lgl a}q_{b\rgl} =
0$, then, to first order,
$$
\D_{\lgl a}v_{b\rgl} = 0 \ ,
$$
which, for $v^{a} \neq 0$, we have recognised to characterise
tilted FLRW models. In view of this result and the need to obtain
{\em consistent\/} Newtonian limits from the exact equations of
General Relativity in a cosmological context, enforcing the PFE by
neglecting the first-order momentum density $q^{a}$ of the matter
in the $(0\alpha )$-EFE seems questionable, even though this term
is certainly very small.

It may be that other approaches, e.g., using `synchronous' gauge
conditions, get around this problem in a satisfactory way. However,
for example Peebles' approach to the CMBR anisotropies, that is
largely based on this gauge, in fact mixes {\em different\/} gauges
--- setting up the CMBR anisotropy investigation in one gauge, and
using another one to establish the PFE relating density
inhomogeneities to the gravitational scalar potential (\,see
Ref. \ct{pee93}, pp500-6\,). This seems a problematic procedure,
and does not appear to solve the issue raised here.

\section{Conclusion}
\l{sec:concl}
Put in loose terms, in this article we have asked the question as
to what extent we can squeeze a generic dust matter source field,
which typically generates non-zero $\tl{\Th}$, $\tl{\sig}_{ab}$,
$\tl{\om}^{a}$, $\tl{E}_{ab}$ and $\tl{H}_{ab}$, into a
Newtonian-like reference frame characterised by the conditions
(\r{zerovor}) - (\r{acc}). The implied invariant physical
requirement that a quasi-Newtonian cosmology $\left(\,{\cal
M},\,{\bf g},\,{\bf n}\,\right)$ excludes gravitational radiation
provides a very strong restriction on the generality of the
resulting cosmological model.
It may turn out that this constraint proves to be sufficiently
strong indeed that no exact spatially inhomogeneous solutions (in
non-comoving description) exist at all. The complexity of the
constraint analysis encountered in section \r{sec:qncons} is
certainly a reflection of the restrictiveness of the
no-gravitational-radiation condition.\footnote{Note that this is
different from other investigations of cosmological models without
gravitational radiation, in particular those with $H_{ab} = 0$
\ct{vanetal97}, because in those cases the
zero-magnetic-Weyl-curvature condition was imposed in the
matter-comoving frame.} Indeed, it appears plausible that after
eliminating the non-Newtonian spacetime curvature source terms
(isotropic and anisotropic pressure) by FLRW-linearisation of the
quasi-Newtonian equations, the gravitational radiation inducing
quantities have been switched off, and the system of dynamical
equations becomes tractable (and possibly solvable). If it could be
shown that non-trivial solutions exist for the linear case but not
the exact equations, linearisation instability of this dynamical
setting would be shown.

Whether or not this is true, our analysis indicates problems
arising in the Newtonian limit procedures in the literature based
on quasi-Newtonian type conditions of the kind we discuss; use of
mixed gauges as in Peebles' discussion \ct{pee93} does not clarify
these issues.

(1) The approximations made in the standard literature, ultimately
leading to the PFE of the Newtonian theory of gravitation, cannot
be regarded as providing an example of a {\em well\/}-defined
(intrinsically consistent) Newtonian limit of the (cosmological)
EFE, as demanded by Ehlers in, e.g., Refs. \ct{ehl81} and
\ct{ehl97}.

(2) If we use a covariant and gauge-invariant approach, this gives
a setting where the ten general relativistic gravitational
potentials reduce effectively to one, as in Newtonian theory (hence
the common use of this approach in linearised studies). However, we
still have unresolved the issue of what determines a suitable
choice for time evolution of the $N(x^{i})$ --- we will get
restricted results if we {\em impose\/} a choice for $\dot{N}$; but
we have been unable to see how the integrability conditions
generically {\em determine\/} the choice of $N(x^{i})$ (which is
what we expected). Thus, we do not have a solution in sight that
satisfactorily resolves the problem. This correponds to the
unresolved question of how one obtains time-evolution for the
Newtonian gravitational scalar potential with realistic
cosmological boundary conditions.

(3) Factoring out the subset of (inhomogeneously perceived)
spatially flat FLRW solutions in the linearised case characterised
by the condition $\D_{\lgl a}v_{b\rgl} = 0 \Rightarrow
\tl{\sig}_{ab} = 0$,\footnote{The spatial rotation
$\epsilon^{abc}\,\D_{b}v_{c} = 0 \Rightarrow \tl{\om}^{a} = 0$ is
of {\em second\/}-order smallness in the first place.} we obtain
extra velocity-dependent terms in the Newtonian-like gravitational
field equations resulting from consistent linearisation of the
relativistic equations, rather than just the PFE that many prefer
to use.

(4) Underlying all this is the difficult issue of coarse graining:
in essence we try to describe {\em two\/} distance scales
(sub-Hubble-radius scales vs cosmological/spatial curvature
relevant scales) by just {\em one\/} set of (mathematically
consistent) dynamical equations. The relation between these
different scales of analysis (and the corresponding effect on the
EFE) needs to be clearly brought out and clarified.

Astrophysically oriented readers may immediately point out that our
use of geometrised physical units disguises the true orders of
magnitudes of the physical effects induced by various terms in our
dynamical equations. For example, upon re-establishing the
fundamental constants $c$ and $8\pi G/c^{2}$ (\,in SI units we have
$8\pi G/c^{4} = 2.076\times 10^{-43}\,s^{2}/(kg\,m)$\,), the
constraint $(C_{1})^{a}$ of Eq. (\r{linc1}) reads
$$
0 = \D^{a}\Th - \sfrac{12\pi G}{c^{2}}\,\rho\,v^{a} \ ,
$$
where $\rho$ denotes the mass density of the matter
constituents. Similarly, non-standard source terms proportional to
$v^{a}$ in each of Eqs. (\r{linc5}) and (\r{delddot}) become quite
small for non-relativistic magnitudes of $v^{a}$. Thus, in
particular cases such as gravitational lensing settings, the extra
terms may be negligible. Such arguments must, however, be pursued
with care; as J L Synge has pointed out (\,see Ref. \ct{syn60},
pp176-8\,), if you too easily drop pressure gradient terms on the
basis of similar arguments, you may deduce that an ocean liner will
sink to the bottom of the sea. The argument needs to be carefully
made in each context, for example, in Zel'dovich-like gravitational
collapse scenarios. What we are pointing out is the necessity when
doing so to look at the consistency of the whole set of equations,
rather than looking at just one or two equations and ignoring the
rest.

\section*{Acknowledgements}
We are grateful to Bruce Bassett, J\"{u}rgen Ehlers and Roy
Maartens for very helpful comments. This work has been supported by
the South African Foundation for Research and Development (FRD) and
the Deutsche Forschungsgemeinschaft (DFG). In parts the computer
algebra package {\tt CLASSI} was employed.

\appendix
\section{Appendix}
\subsection{Commutation relations}
\l{subsec:com}
The following commutation relations for $1+3$ covariantly defined
scalars $f$ and vectors $V^{a}$ have been frequently used within
our work:
\bea
\l{3sclcom}
h^{b}\!_{a}\,(\D_{b}f)\,\dot{}
& = & [\ \D_{a} + \ndot_{a}\ ]\,(\dot{f})
- \sfrac{1}{3}\,\Th\,\D_{a}f \\ \nonumber \\
\l{vec3ricid}
\D_{[a}\D_{b]}V_{c} & = & E_{c[a}\,V_{b]}
+ \sfrac{1}{2}\,\mu\,v_{c}\,v_{[a}\,V_{b]}
+ h_{c[a}\,E_{b]d}\,V^{d}
+ \sfrac{1}{2}\,(v_{d}V^{d})\,\mu\,h_{c[a}\,v_{b]}
\\ \nonumber \\
& & \hsp5 + \ \sfrac{1}{3}\,[\ (1-v^{2})\,\mu
- \sfrac{1}{3}\,\Th^{2}\ ]\ h_{c[a}\,V_{b]} \\ \nonumber \\
\l{convec3ricid}
\D_{[a}\D_{b]}V^{b} & = & -\,\sfrac{1}{2}\,E_{ab}\,V^{b}
- \sfrac{1}{4}\,(v_{b}V^{b})\,\mu\,v_{a}
- \sfrac{1}{3}\,[\ (1-\sfrac{1}{4}v^{2})\,\mu
- \sfrac{1}{3}\,\Th^{2}\ ]\ V_{a} \ .
\eea
Recall that in general the commutation relation for spatial
derivative operators acting on $f$ is given by $\D_{[a}\D_{b]}f =
\eps_{abc}\,\om^{c}\,\dot{f}$ (\,cf. Ref. \ct{elletal90}\,).
However, as for quasi-Newtonian cosmologies $\om^{a}({\bf n}) = 0$, 
within the ${\cal T}$:
$\left\{t=\mbox{const}\right\}$ we have
\be
\l{zerocom}
\D_{[a}\D_{b]}f = 0 \ .
\ee
%

\subsection{Spatial derivatives of $v^{a}$}
The irreducible decomposition of the purely spatial covariant
derivative of the peculiar velocity $v^{a}$ is given by
\be
\D_{a}v_{b} = \D_{\lgl a}v_{b\rgl} + \sfrac{1}{3}\,(\D_{c}v^{c})\,
h_{ab} + \eps_{abc}\,[\ \sfrac{1}{2}\,\eps^{cde}\,\D_{d}v_{e}\ ]
\ . 
\ee
Now defining
\be
p_{ab} := h_{ab} - v_{a}/v\,v_{b}/v \ , \hsp5
q_{ab} := v_{a}/v\,v_{b}/v - \sfrac{1}{2}\,p_{ab} \ , \hsp5
s_{ab} := \eps_{abc}\,v^{c}/v \ ,
\ee
individual derivative terms can be expressed by
\bea
\D_{\lgl a}v_{b\rgl} & = & [\ 2\,v_{\lgl a}/v\,p^{\lgl c}{}_{b\rgl}
\,v^{d\rgl}/v + \sfrac{2}{3}\,q_{ab}\,q^{cd}
+ p^{c}\!_{\lgl a}\,p^{d}\!_{b\rgl}\ ]\,(\D_{c}v_{d}) \\
\D_{a}v^{a} & = & \sfrac{1}{2}\,v^{a}/v^{2}\,\D_{a}v^{2}
+ p^{a}\!_{b}\D_{a}v^{b} \\
\l{pvrot2}
\eps^{abc}\,\D_{b}v_{c} & = & v^{a}/v\,(s^{bc}\,\D_{b}v_{c})
+ 2\,s^{a[b}\,v^{c]}/v\,(\D_{b}v_{c}) \\
\D_{a}v^{2} & = & v_{a}/v\,(v^{b}/v\,\D_{b}v^{2})
+ p^{b}\!_{a}\,\D_{b}v^{2} \\
v^{b}\D_{b}v^{a} & = & \sfrac{1}{2}\,v^{a}/v\,(v^{b}/v\,
\D_{b}v^{2}) + p^{a}\!_{b}\,v^{c}\D_{c}v^{b} \ .
\eea
Remember that the spatial rotation of $v^{a}$ as given in
Eq. (\r{pvrot2}) is algebraically constrained through
Eq. (\r{pvrot}).

\subsection{Kinematical and Weyl curvature variables relating to
the dust matter flow}
\l{subsec:matuder}
The dust matter flow we discuss is moving with average 4-velocity
${\bf \tl{u}}$; we have $\dot{\tl{u}}{}^{a} = 0$. In deriving the
following set of relations we made use of Eqs. (\r{tlu}),
(\r{covvdot}), (\r{pvrot}) and (\r{pvdec}):
\bea
\l{tlusig}
\tl{\sig}_{ab}(v) & = & \gam\,[\ \D_{\lgl a}v_{b\rgl}
+ \sfrac{2}{3}\,\Th\,n_{(a}\,v_{b)} + \sfrac{1}{3}\,\Th\,
v^{2}\,(n_{a}\,n_{b}+\sfrac{1}{3}\,h_{ab})
+ n_{(a}\,v^{c}\D_{|c|}v_{b)}\ ] \nonumber \\
& & \hsp5 + \ \gam^{3}\ [\ \sfrac{1}{2}\,(n_{(a}+v_{(a})\,
\D_{b)}v^{2} + n_{(a}\,(n_{b)}+v_{b)})\,(v^{c}v^{d}\D_{c}v_{d})
\ ] \nonumber \\
& & \hsp5 - \ \sfrac{1}{3}\,\gam^{3}\ [\ (1-\sfrac{1}{3}v^{2})\,
\Th + \D_{c}v^{c}\ ]\,[\ 2\,n_{(a}\,v_{b)} + v^{2}\,n_{a}\,n_{b}
+ v_{a}\,v_{b}\ ]
\\ \nonumber \\
\l{tluth}
\tl{\Th}(v) & = & \gam\,[\ (1-\sfrac{1}{3}v^{2})\,\Th
+ \D_{a}v^{a}\ ]
\\ \nonumber \\
\l{tluom}
\tl{\om}_{ab}(v) & = & \gam\,[\ \D_{[a}v_{b]}
+ n_{[a}\,v^{c}\D_{|c|}v_{b]}
- \sfrac{1}{2}\,\gam^{2}\,(n_{[a}+v_{[a})\,\D_{b]}v^{2}
+ \gam^{2}\,n_{[a}\,v_{b]}\,(v^{c}v^{d}\D_{c}v_{d})\ ] \ .
\eea
As usual, the vorticity vector $\tl{\om}^{a}$ is given by
\be
\tl{\om}^{a} = \sfrac{1}{2}\,\tl{\eps}^{abc}\,\tl{\om}_{bc} \ .
\ee
In the non-comoving FLRW-linearised limit the kinematical variables
of the dust matter flow are given by
\bea
\l{lintlsig}
\tl{\sig}_{ab} & = & \D_{\lgl a}v_{b\rgl} \\
\l{lintlth}
\tl{\Th} & = & \Th + \D_{a}v^{a} \\
\l{lintlom}
\tl{\om}_{ab} & = & 0 \ ;
\eea
recall that by Eq. (\r{pvrot}) $\D_{[a}v_{b]}$ is of second-order
smallness.

The Weyl curvature variables with respect to the dust matter flow
${\bf \tl{u}}$ can be obtained from
\bea
\l{tluele}
\tl{E}_{ab}({\bf \tl{u}}) & = & -\,\dot{\tl{\sig}}_{\lgl ab\rgl}
- \sfrac{2}{3}\,\tl{\Th}\,\tl{\sig}_{ab}
- \tl{\sig}^{c}\!_{\lgl a}\,\tl{\sig}_{b\rgl c}
- \tl{\om}_{\lgl a}\,\tl{\om}_{b\rgl} \\
& = & \gam^{2}\,[\ (1+v^{2})\,E_{ab} + 2\,(n_{(a}-v_{(a})\,
E_{b)c}\,v^{c} + (E_{cd}v^{c}v^{d})\,(n_{a}\,n_{b}
+ v_{a}/v\,v_{b}/v + p_{ab})\ ]
\\ \nonumber \\
\l{tlumag}
\tl{H}_{ab}({\bf \tl{u}}) & = & -\,\tl{\D}_{\lgl a}\tl{\om}_{b\rgl}
+ \tl{\eps}_{cd\lgl a}\,\tl{\D}^{c}\tl{\sig}^{d}\!_{b\rgl} \\
& = & 2\,\gam^{2}\,\eps_{cd\lgl a}\,[\ n_{b\rgl}\,(E^{c}\!_{e}
v^{e})  + E^{c}\!_{b\rgl}\ ]\,v^{d} \ .
\eea
%

\subsection{$1+3$ orthonormal frame dynamical equations}
\l{subsec:13onfqn}
For a full consistency analysis, we need the dynamical equations
describing quasi-Newtonian cosmologies with dust matter source
field expressed relative to a $1+3$ orthonormal frame $\{\,{\bf
e}_{a}\,\}$ (\,see, e.g., Refs. \ct{ell67} and
\ct{hveugg97}\,). After aligning the timelike frame vector field
${\bf e}_{0}$ with our reference congruence ${\bf n}$, i.e., ${\bf
e}_{0} = {\bf n}$, one can choose a spatial frame $\{\,{\bf
e}_{\alpha}\,\}$ according to convenience. Making use of this
freedom, the following set of quasi-Newtonian relations, which we
derive from the general equations given in Ref. \ct{hveugg97} by
simply deleting terms, can be further simplified.\enl

\noindent
{\em Commutation relations\/}:
\bea
\l{onfcomts}
\left[\,{\bf e}_{0}, {\bf e}_{\alpha}\,\right] & = &
\ndot_{\alpha}\,{\bf e}_{0} - \left[\ \sfrac{1}{3}\,\Th\,
\delta^{\beta}\!_{\alpha} - \epsilon^{\beta}\!_{\alpha\gamma}\,
\Omega^{\gamma}\ \right]\ {\bf e}_{\beta}
\\ 
\l{onfcomss}
\left[\,{\bf e}_{\alpha}, {\bf e}_{\beta}\,\right] & = &
\left[\ 2\,a_{[\alpha}\,\delta^{\gamma}\!_{\beta]}
+ \epsilon_{\alpha\beta\delta}\,n^{\delta\gamma}\ \right]
\ {\bf e}_{\gamma} \ .
\eea
\noindent
{\em Einstein field equations\/}:
\bea
\l{onfthdot}
{\bf e}_{0}(\Th) & = & - \,\sfrac{1}{3}\,\Th^{2}
+ ({\bf e}_{\alpha} + \ndot_{\alpha} - 2\,a_{\alpha})\,
(\ndot^{\alpha}) - \sfrac{1}{2}\,(1+v^{2})\,\mu \\
\l{onfsigdot}
0 & = & (\delta^{\gamma\lgl\alpha}\,{\bf e}_{\gamma}
+ \ndot^{\lgl\alpha} + a^{\lgl\alpha})\,(\ndot^{\beta\rgl})
+ \mu\,v^{\lgl\alpha}\,v^{\beta\rgl} - {}^{*}\!S^{\alpha\beta}
- \epsilon^{\gamma\delta\lgl\alpha}\,n^{\beta\rgl}\!_{\gamma}\,
\ndot_{\delta} \\
\l{onffried}
0 & = & {}^*\!R + \sfrac{2}{3}\,\Th - 2\,\mu \ = \ (C_{G}) \\
\l{onfdivsig}
0 & = & \delta^{\alpha\beta}\,{\bf e}_{\beta}(\Th)
- \sfrac{3}{2}\,\mu\,v^{\alpha} \ = \ (C_{1})^{\alpha} \ ,
\eea
where
\bea
\l{onftf3ric}
{}^{*}\!S_{\alpha\beta} & := & {\bf e}_{\lgl\alpha}(a_{\beta\rgl})
+ b_{\lgl\alpha\beta\rgl}
- \epsilon^{\gamma\delta}\!_{\lgl\alpha}\,({\bf e}_{|\gamma|}
- 2\,a_{|\gamma|})\,(n_{\beta\rgl\delta}) \\
\l{onf3rscl}
{}^*\!R & := &  2\,(2\,{\bf e}_{\alpha} - 3\,a_{\alpha})\,
(a^{\alpha}) - {\sfrac{1}{2}}\,b^{\alpha}\!_{\alpha} \\
b_{\alpha\beta} & := & 2\,n_{\alpha\gamma}\,n^{\gamma}\!_{\beta}
- n^{\gamma}\!_{\gamma}\,n_{\alpha\beta} \ .
\eea
\noindent
{\em Jacobi identities\/}:
\bea
\l{jac1}
{\bf e}_{0}(a^{\alpha}) & = & - \,\sfrac{1}{3}\,
(\delta^{\alpha\beta}\,{\bf e}_{\beta} + \ndot^{\alpha}
+ a^{\alpha})\,(\Th) + {\sfrac{1}{2}}\,\epsilon^{\alpha\beta\gamma}
\,({\bf e}_{\beta} + \ndot_{\beta} - 2\,a_{\beta})\,
(\Omega_{\gamma}) \\
\l{jac2}
{\bf e}_{0}(n^{\alpha\beta}) & = & - \,\sfrac{1}{3}\,\Th\,
n^{\alpha\beta} + (\delta^{\gamma(\alpha}\,{\bf e}_{\gamma}
+ \ndot^{(\alpha})\,(\Omega^{\beta)})
- \delta^{\alpha\beta}\,({\bf e}_{\gamma}
+ \ndot_{\gamma})\,(\Omega^{\gamma})
- 2\,\epsilon^{\gamma\delta(\alpha}\,n^{\beta)}\!_{\gamma}\,
\Omega_{\delta} \\
\l{jac3}
0 & = & \epsilon^{\alpha\beta\gamma}\,({\bf e}_{\beta} - a_{\beta})\,
(\ndot_{\gamma}) - n^{\alpha}\!_{\beta}\,\ndot^{\beta}
\ = \ (C_{2})^{\alpha} \\
\l{jac4}
0 & = & ({\bf e}_{\beta} - 2\,a_{\beta})\,(n^{\alpha\beta})
+ \epsilon^{\alpha\beta\gamma}\,{\bf e}_{\beta}(a_{\gamma}) \ .
\eea
\noindent
{\em Electric Weyl curvature\/}:
\bea
\l{onfele1}
E_{\alpha\beta} & = & ({\bf e}_{\lgl\alpha}
+ \ndot_{\lgl\alpha} + a_{\lgl\alpha})\,(\ndot_{\beta\rgl})
+ \sfrac{1}{2}\,\mu\,v_{\lgl\alpha}\,v_{\beta\rgl}
- \epsilon^{\gamma\delta}\!_{\lgl\alpha}\,n_{\beta\rgl\gamma}\,
\ndot_{\delta} + (C_{3})_{\alpha\beta} \\
& = & {}^{*}\!S_{\alpha\beta}
- \sfrac{1}{2}\,\mu\,v_{\lgl\alpha}\,v_{\beta\rgl}
- (C_{G})_{\alpha\beta} \ .
\eea
\noindent
{\em Contracted Bianchi identities\/}:
\bea
\l{onfmudot}
{\bf e}_{0}(\mu) & = & -\,(1+\sfrac{1}{3}v^{2})\,\Th\,\mu
- v^{\alpha}\,{\bf e}_{\alpha}(\mu) - \mu\,({\bf e}_{\alpha}
+ 2\,\ndot_{\alpha} - 2\,a_{\alpha})(v^{\alpha}) \\
\l{onfqdot}
{\bf e}_{0}(v^{\alpha}) & = & -\,\sfrac{1}{3}\,(1-v^{2})\,\Th\,
v^{\alpha} - v^{\beta}\,({\bf e}_{\beta} - \ndot_{\beta}
- a_{\beta})\,(v^{\alpha}) - v^{2}\,a^{\alpha} - \ndot^{\alpha}
\nonumber \\
& & \hsp5 + \ \epsilon^{\alpha\beta\gamma}\,
\left[\ n_{\beta\delta}\,v^{\delta}\,v_{\gamma}
+ \Omega_{\beta}\,v_{\gamma}\ \right] \ .
\eea
%

\subsection{Line element and linearised Einstein field equations
for scalar-perturbed spatially flat FLRW dust}
\l{subsec:flrw0p}
Many perturbation calculations use the following quasi-Newtonian
line element, where $\eta$ denotes a dimensionless conformal time
coordinate and $x$, $y$ and $z$ are Cartesian coordinates comoving
with ${\bf n}$: \enl

\noindent
{\em Infinitesimal line element\/}:
\be
\l{flrw0pds2}
ds^{2} = S^{2}\,[\ -\,(1+2\Phi)\,d\eta^{2} + (1-2\Psi)\,
(\,dx^{2} + dy^{2} + dz^{2}\,)\ ] \ .
\ee
The associated commutation functions and components of the Einstein
curvature tensor, as well as the first-order components of the
energy-momentum-stress tensor are:\enl

\noindent
{\em Commutation functions\/}:
\be
\ndot_{\alpha} = \frac{1}{S}\,\Phi_{|\alpha} \hspace{1cm}
\Th = \frac{3}{S}\left[\ (1-\Phi)\,\frac{S_{|\eta}}{S}
- \Psi_{|\eta}\ \right] \hspace{1cm}
a_{\alpha} = \frac{1}{S}\,\Psi_{|\alpha} \ .
\ee
$0 = \sig_{\alpha\beta} = \om^{\alpha} = \Omega^{\alpha}
= n_{\alpha\beta}$.\enl

\noindent
{\em Einstein curvature tensor\/}:
\bea
\l{ling00}
G_{00} & = & \frac{1}{S^{2}}\,\left[\ 3\,(1-2\Phi)
\left(\frac{S_{|\eta}}{S}\right)^{2} - 6\,\frac{S_{\eta}}{S}
\,\Psi_{|\eta} + 2\,\Psi^{|\alpha}\!_{|\alpha}\ \right] \\
\l{ling0a}
G_{0\alpha} & = & \frac{2}{S^{2}}\,\left[\ \frac{S_{|\eta}}{S}\,
\Phi_{|\alpha} + \Psi_{|\eta\alpha}\ \right] \\
\l{lingaa}
G_{\alpha\alpha} & = & -\,\frac{1}{S^{2}}\,\left[\ 2\,(1-2\Phi)
\,\frac{S_{|\eta\eta}}{S} - (1-2\Phi)
\left(\frac{S_{|\eta}}{S}\right)^{2} \right.
\hspace{1cm} \mbox{no summation,}\ \alpha \neq \beta \neq \gam
\nonumber \\
& & \hspace{2cm} \left. - \ 2\,\frac{S_{|\eta}}{S}\,
(\Phi+2\Psi)_{|\eta} - 2\,\Psi_{|\eta\eta}
- (\Phi-\Psi)_{|\beta\beta} - (\Phi-\Psi)_{|\gam\gam}\ \right] \\
\l{lingab}
G_{\alpha\beta} & = & -\,\frac{1}{S^{2}}
\,(\Phi-\Psi)_{|\alpha\beta} \hspace{1cm} \alpha \neq \beta \ .
\eea
\noindent
{\em Energy-momentum-stress tensor\/}:
\be
T_{00} = \mu \hspace{1cm}
T_{0\alpha} = \mu\,v_{\alpha} \hspace{1cm}
T_{\alpha\beta} = 0 \ .
\ee
%




\begin{thebibliography}{99}

\bibitem{abretal97}
Abrahams A, Anderson A, Choquet-Bruhat Y and York Jr J W 1997
Geometrical hyperbolic systems for general relativity and gauge
theories \cqg {\bf 14} A9

\bibitem{adm62}
Arnowitt R, Deser S and Misner C W 1962 The dynamics of general
relativity {\em Gravitation\/} ed L Witten (New York: Wiley)

\bibitem{bar80}
Bardeen J M 1980 Gauge-invariant cosmological perturbations \prd
{\bf 22} 1882

\bibitem{ber92}
Bertschinger E 1992 Large-scale structures and motions: Linear
theory and statistics {\em Current Topics in Astrofundamental
Physics\/} ({\em International School of Astrophysics `D
Chalonge'\/}) eds N Sanchez and A Zichichi (Singapore: World
Scientific)

\bibitem{bonetal95}
Bona C, Mass\'{o} J, Seidel E and Stela J 1995 A new formalism for
numerical relativity \prl {\bf 75} 600

\bibitem{bruetal92}
Bruni M, Dunsby P K S and Ellis G F R 1992 Cosmological
perturbations and the physical meaning of gauge-invariant
variables {\em Astrophys. J.\/} {\bf 395} 34

\bibitem{ehl61}
Ehlers J 1961 Beitr\"{a}ge zur re\-la\-ti\-vis\-ti\-schen
Me\-cha\-nik kon\-ti\-nu\-ier\-li\-cher Me\-di\-en {\em
Akad. Wiss. Lit. Mainz, Abhandl. Math.-Nat. Kl.\/} {\bf 11} 792
\enl
Reprinted in 1993 \grg {\bf 25} 1225

\bibitem{ehl81}
Ehlers J 1981 \"{U}ber den Newtonschen Grenzwert der Einsteinschen
Gravitationstheorie {\em Grundlagenprobleme der modernen Physik\/}
ed J Nitsch, J Pfarr and E-W Stachow (Mannheim: Bibliographisches
Institut)

\bibitem{ehl97}
Ehlers J 1997 Examples of Newtonian limits of relativistic
spacetimes \cqg {\bf 14} A119

\bibitem{ehlsch93}
Ehlers J and Schneider P 1993 Gravitational lensing {\em General
Relativity\/} ({\em Proceedings of the 13th International
Conference\/}) eds R J Gleiser, C N Kozameh and O M Moreschi
(Bristol: Institute of Physics Publishing)

\bibitem{ell67}
Ellis G F R 1967 Dynamics of pressure-free matter in general
relativity {\em J. Math. Phys.\/} {\bf 8} 1171

\bibitem{ell71}
Ellis G F R 1971 Relativistic cosmology {\em General
Relativity and Cosmology\/} ({\em Proc. 47th Enrico Fermi
Summer School\/}) ed R K Sachs (New York: Academic Press)

\bibitem{ell84}
Ellis G F R 1984 Relativistic cosmology: Its nature, aims and
problems {\em General Relativity and Gravitation\/} ({\em Invited
Papers and Discussion Reports of the 10th International
Conference\/}) ed B Bertotti, F de Felice and A Pascolini
(Dordrecht: Reidel)

\bibitem{ell87}
Ellis G F R 1987 Standard cosmology {\em Vth Brazilian School on
Cosmology and Gravitation\/} ed M Novello (Singapore: World
Scientific)

\bibitem{ell97}
Ellis G F R 1997 Cosmological models from a covariant viewpoint
{\em Gravitational Cosmology (Proc. ICGC95 Conf. at IUCCA
(Pune))\/} ed S Dhurandhar and T Padmanabhan (Dordrecht: Kluwer)

\bibitem{ellsch86}
Ellis G F R and Schreiber G 1986 Observational and dynamic
properties of small universes {\em Phys. Lett.\/} {\bf 115A} 97

\bibitem{ellbru89}
Ellis G F R and Bruni M 1989 Covariant and gauge--invariant
approach to cosmological density fluctuations \prd {\bf 40} 1804

\bibitem{elletal90}
Ellis G F R, Bruni M and Hwang J 1990 Density-gradient--vorticity
relation in perfect-fluid Robertson--Walker perturbations \prd {\bf
42} 1035

\bibitem{ellmat95}
Ellis G F R and Matravers D R 1995 General covariance in general 
relativity \grg {\bf 27} 777

\bibitem{ellhve98}
Ellis G F R and van Elst H 1998 Consistency of inhomogeneous
solutions and the nature of the Newtonian limit {\em Black Holes
and High Energy Astrophysics\/} ({\em 49th Yamada Conference\/}) To
Appear

\bibitem{hve96}
van Elst H 1996 Extensions and applications of $1+3$
decomposition methods in general relativistic cosmological
modelling {\em PhD thesis\/} University of London\enl
Available at {\tt http://shiva.mth.uct.ac.za/$\sim$henk/abstr.html}

\bibitem{hveell96}
van Elst H and Ellis G F R 1996 The covariant approach to LRS
perfect fluid spacetime geometries \cqg {\bf 13} 1099

\bibitem{hveugg97}
van Elst H and Uggla C 1997 General relativistic $1+3$ orthonormal
frame approach \cqg {\bf 14} 2673

\bibitem{vanetal97}
van Elst H, Uggla C, Lesame W M, Ellis G F R and Maartens R 1997
Integrability of irrotational silent cosmological models \cqg {\bf
14} 1151

\bibitem{fri98}
Friedrich H 1998 Evolution equations for gravitating ideal fluid
bodies in general relativity \prd {\bf 57} 2317

\bibitem{hecsch56}
Heckmann O and Sch\"{u}cking E 1956 Bemerkungen zur Newtonschen
Kosmologie. II. {\em Zeits. f. Astroph.\/} {\bf 40} 81

\bibitem{holwal97}
Holz D E and Wald R M 1998 A new method for determining cumulative
gravitational lensing effects in inhomogeneous universes \prd {\bf
58} 063501

\bibitem{kinell73}
King A R and Ellis G F R 1973 Tilted homogeneous cosmological
models {\em Commun. Math. Phys.\/} {\bf 31} 209

\bibitem{ksmh80}
Kramer D, Stephani H, MacCallum M A H and Herlt E 1980 {\em Exact
Solutions of Einstein's Field Equations\/} (Berlin: VEB Dt. Verlag
d. Wissenschaften)

\bibitem{maa97}
Maartens R 1997 Linearisation instability of gravity waves? \prd
{\bf 55} 463

\bibitem{maa98}
Maartens R 1998 Covariant velocity and density perturbations in
quasi-Newtonian cosmologies \prd At Press

\bibitem{maaetal98}
Maartens R, Lesame W M and Ellis G F R 1998 Newtonian-like and
anti-Newtonian universes \cqg {\bf 15} 1005

\bibitem{mtw73}
Misner C W, Thorne K S and Wheeler J A 1973 {\em Gravitation\/}
(New York: Freeman \& Co)

\bibitem{muketal92}
Mukhanov V F, Feldman H A and Brandenberger R H 1992 Theory of
cosmological perturbations {\em Phys. Rep.\/} {\bf 215} 203

\bibitem{pee69}
Peebles P J E 1969 Cosmology for everyphysicist {\em
Am. J. Phys.\/} {\bf 37} 410

\bibitem{pee80}
Peebles P J E 1980 {\em The Large-Scale Structure of the
Universe\/} (Princeton: Princeton University Press)

\bibitem{pee93}
Peebles P J E 1993 {\em Principles of Physical Cosmology\/}
(Princeton: Princeton University Press)

\bibitem{seietal94}
Seitz S, Schneider P and Ehlers J 1994 Light propagation in
arbitrary spacetimes and the gravitational lens approximation \cqg
{\bf 11} 2345

\bibitem{syn60}
Synge J L 1960 {\em Relativity: The General Theory\/} (Amsterdam:
North-Holland)

\bibitem{treell71}
Treciokas R and Ellis G F R 1971 Isotropic solutions of the
Einstein--Boltzmann equations {\em Commun. Math. Phys.\/} {\bf 23}
1

\bibitem{vel97}
Velden T 1997 Dynamics of pressure-free matter in general
relativity {\em Diplomarbeit\/} Universit\"{a}t
Bielefeld/Albert--Einstein--Institut Potsdam

\bibitem{wai79}
Wainwright J 1979 A classification scheme for non-rotating
inhomogeneous cosmologies {\em J. Phys. A: Math. Gen.\/} {\bf 12}
2015

\bibitem{whi35}
Whittaker E T 1935 On Gauss' Theorem and the concept of mass
in general relativity {\em Proc. R. Soc. London A\/} {\bf 149}
384

\bibitem{wal84}
Wald R M 1984 {\em General Relativity\/} (Chicago: University of
Chicago Press)

\bibitem{yor79}
York Jr J W 1979 Kinematics and dynamics of general relativity {\em
Sources of Gravitational Radiation, Proceedings of the Battelle
Seattle Workshop\/} ed L L Smarr (Cambridge: Cambridge University
Press)

\begin{center}
* * *
\end{center}

\end{thebibliography}
\end{document}